\documentclass[a4paper,fleqn]{cas-sc}

\ExplSyntaxOn
\cs_set:Npn \__first_footerline:
  {
    \group_begin:
    \small\sffamily
    \ifnum\theblind>0\relax\else
      \__short_authors:
    \fi
    \group_end:
  }
\ExplSyntaxOff

\ifdefined\pdfcompresslevel
\fi

\usepackage[numbers,sort&compress]{natbib}

\usepackage{amsmath,amssymb}
\usepackage{booktabs,tabularx}
\usepackage{multirow}
\usepackage{placeins}


\graphicspath{{figs/}}

\def\tsc#1{\csdef{#1}{\textsc{\lowercase{#1}}\xspace}}
\tsc{WGM}
\tsc{QE}

\begin{document}
\let\WriteBookmarks\relax

\shorttitle{}

\shortauthors{Liu and Yao}

\title [mode = title]{Learning structural response operators across varying geometries and loads with global and local geometry encoding}


\author[1]{Xin Liu}

\author[1]{Wuwen Yao}[orcid=0000-0002-5842-5373]
\cormark[1]
\ead{geotech.yaowuwen@gmail.com}

\affiliation[1]{organization={School of Civil and Environmental Engineering, Changsha University of Science and Technology},
            city={Changsha},
            country={China}}

\cortext[1]{Corresponding Author: Wuwen Yao}

\begin{abstract}
Predicting mechanical responses across complex geometries and varying loading conditions is central to simulation-driven engineering, yet capturing the influence of geometric boundaries remains challenging for data-driven models. Here we propose a Load- and Geometry-adaptive Deep Operator Network (LG-DON) that integrates global geometric representations with local, boundary-aware information to predict structural responses. By distinguishing distances to different boundary categories rather than relying on the nearest boundary alone, the framework captures spatially varying geometric influences that conventional distance-based representations may overlook. Across benchmark problems involving diverse non-convex geometries, LG-DON reduces the mean relative $L_2$ error from 12.17\% for a modified MIONet baseline to 7.41\%. Controlled ablation studies further demonstrate the advantages of category-resolved boundary distances and show that combining global and local geometric representations outperforms either representation alone in 15 of 18 evaluated configurations. These findings highlight the value of integrating complementary geometric information for accurate operator learning across complex structural domains.
\end{abstract}




\maketitle

\section{Introduction}\label{sec:introduction}

Structural design, optimization and uncertainty analysis require repeated predictions of displacement and stress as geometry and boundary excitations change. Finite-element methods provide reference solutions, but changing a domain may require geometry reconstruction and remeshing, followed by another numerical solve. A surrogate for repeated evaluations must account for both the domain and the functions prescribed along its boundaries. This requirement is clear in structures with irregular holes. Their positions and contours affect the response to a given excitation, while different excitations produce different responses on the same geometry. Such joint variation motivates learning families of structural response operators rather than a map restricted to one domain or loading condition \cite{wang2026lgdon01}.

Conventional neural surrogates often map finite-dimensional inputs to response vectors tied to a prescribed discretization \cite{karniadakis2021lgdon02,li2021lgdon03,raissi2019lgdon04}. Neural operators instead approximate mappings between function spaces \cite{kovachki2023lgdon05}. DeepONet combines an input-function branch with a coordinate-dependent trunk, drawing on operator approximation results \cite{chen1995lgdon06,lu2019lgdon07}. Fourier and graph neural operators represent spatial interactions in other ways \cite{li2020lgdon08,li2020lgdon09,li2020lgdon10}, while physics-informed variants incorporate governing equations or physical constraints during training \cite{wang2021lgdon11,li2024lgdon12}. Multiple-input operator learning is relevant when distinct boundary functions jointly determine a solution. MIONet represents such mappings through separate branches and tensor-product interactions \cite{jin2022lgdon13}. In structural mechanics, LA-PIDON separately encodes traction, prescribed displacement and material parameters \cite{liu2025lgdon14}. These approaches accommodate varying physical inputs, but changing geometry raises a further question: how should the domain be represented to the predictor?

Geometry-adaptive methods differ in how domain information enters the computation. PhyGeoNet, Geo-FNO and DIMON transform physical domains into shared computational or reference spaces \cite{gao2021lgdon15,li2023lgdon16,yin2024lgdon17}. PI-GANO pools sampled geometry points \cite{zhong2025lgdon19}, whereas Geom-DeepONet supplies a signed distance function (SDF) at query locations \cite{he2024lgdon20}. GINO and Point-DeepONet combine point-derived geometry with distance information \cite{li2023lgdon21,park2026lgdon22}. GNOT uses attention to handle irregular meshes and multiple inputs \cite{hao2023lgdon23}. GA-VINO combines geometry-aware attention with variational training \cite{wang2026lgdon24}, while CATO learns coordinates for geometry-adaptive attention \cite{cheng2026lgdon25}. DAFNO \cite{liu2023lgdon18} incorporates a domain indicator into operator layers, and GINOT \cite{liu2026lgdon26} connects encoded surface points to queries through attention. Table~\ref{tab:geometry_approaches} groups these approaches by the operation performed on geometry. A model can use more than one of these operations.

\begin{table}[pos=!htbp]
\centering
\caption{Representative approaches to handling variable geometry in learned PDE models.}
\label{tab:geometry_approaches}

\small
\setlength{\tabcolsep}{4pt}
\renewcommand{\arraystretch}{1.23}

\begin{tabularx}{\linewidth}{@{}>{\raggedright\arraybackslash}p{0.265\linewidth}>{\raggedright\arraybackslash}X>{\raggedright\arraybackslash}p{0.25\linewidth}@{}}

\toprule

Geometry-handling approach
&
Operation on geometry
&
Representative models
\\

\midrule

Domain transformation
&
Map the physical domain, its coordinates or its fields to a shared computational or reference space
&
PhyGeoNet \cite{gao2021lgdon15}\textsuperscript{*}, Geo-FNO \cite{li2023lgdon16}, DIMON \cite{yin2024lgdon17}
\\
\addlinespace[2pt]

Domain-indicator conditioning
&
Use an indicator of the physical domain to modify computation on an enclosing grid
&
DAFNO \cite{liu2023lgdon18}
\\
\addlinespace[2pt]

Sample-level geometry encoding
&
Aggregate sampled geometric points into a feature shared across query locations
&
PI-GANO \cite{zhong2025lgdon19}
\\
\addlinespace[2pt]

Query-level distance conditioning
&
Supply an SDF or boundary distance at each prediction location
&
Geom-DeepONet \cite{he2024lgdon20}
\\
\addlinespace[2pt]

Combined point and distance representations
&
Use point-derived geometry together with a spatial distance field
&
GINO \cite{li2023lgdon21}, Point-DeepONet \cite{park2026lgdon22};
\\
\addlinespace[2pt]

Attention-based geometric interaction
&
Relate irregular spatial features, surface points or learned geometric coordinates to prediction locations
&
GNOT \cite{hao2023lgdon23}, GA-VINO \cite{wang2026lgdon24}, CATO \cite{cheng2026lgdon25}, GINOT \cite{liu2026lgdon26}
\\

\bottomrule
\end{tabularx}

\par\vspace{2pt}

{\raggedright\footnotesize
\textsuperscript{*} PhyGeoNet is included as an adjacent geometry-adaptive PDE solver, not as a neural operator.
\par}

\end{table}

Point-set encoders based on shared mappings and permutation-invariant aggregation produce a sample-level geometry feature \cite{qi2017lgdon27,zaheer2017lgdon28}. The same feature is available at every query location, but it does not explicitly give that query's distance to each boundary category. An SDF, by contrast, varies with spatial position \cite{park2019lgdon29}; Geom-DeepONet uses a scalar SDF alongside query coordinates \cite{he2024lgdon20}. In a structure with several boundary categories, a single nearest-boundary distance does not retain the distance to each category. Category-resolved channels preserve these separate relationships, whereas a pooled point-set feature represents the overall configuration.

A finite collection of query-point distances does not generally identify the complete geometry. Likewise, pooling does not explicitly retain every relationship between a query and its boundaries. Previous models show that point-derived and distance-based information can be combined \cite{he2024lgdon20,li2023lgdon21,park2026lgdon22}. The question addressed here is what each representation contributes when the other is already available. We therefore compare point-set selections and boundary-distance configurations both separately and in matched combinations. Evaluating structures with changing internal and external boundaries tests whether those contributions depend on the type of geometric variation.

We address this question with the Load- and Geometry-adaptive Deep Operator Network (LG-DON). Separate branches encode traction and prescribed displacement, while discrete geometry encoding (DG) and query-dependent boundary distances condition the geometry pathway. The distance inputs used here are nonnegative material-side magnitudes, rather than a complete signed field over the ambient space. We evaluate the model on a two-hole plate, a single-hole cantilever and an L-shaped bracket, covering internal- and external-boundary variation. The evaluation includes a modified-MIONet comparison, separate load-amplitude and geometry-parameter extrapolation tests, independently trained discretization studies, and matched geometry-encoding ablations. These tests assess prediction within the studied geometry families and distinguish the contributions of the individual representations from those of their combination.

\section{Methodology}\label{sec:methodology}

\subsection{Problem Statement}\label{subsec:problem_statement}

Structural design, optimization, and uncertainty quantification often require repeated evaluations of structural responses under different loading conditions and geometric configurations. Neural operators trained for varying loads on a fixed domain, or for varying geometries under fixed loading conditions, approximate only restricted forms of the full structural response operator. In this work, we consider simultaneous variations in loading conditions and structural geometry and formulate the task as learning a load--geometry-dependent structural response operator over a family of varying physical domains.

Let $\mathfrak{D}$ denote the family of structural domains considered in this study. For the $i$-th sample, the corresponding physical domain is denoted by
\begin{equation}\label{eq:domain}
\Omega^{(i)} \in \mathfrak{D}.
\end{equation}
Under the assumption of small-deformation linear elasticity, the displacement vector is defined as
\begin{equation}\label{eq:displacement}
\mathbf{u}^{(i)}(\mathbf{x}) =
\begin{bmatrix}
u_x^{(i)}(\mathbf{x}) & u_y^{(i)}(\mathbf{x})
\end{bmatrix}^{T},
\qquad \mathbf{x}\in\Omega^{(i)}.
\end{equation}
The static equilibrium equation in the interior of the structure is
\begin{equation}\label{eq:equilibrium}
\nabla\cdot\boldsymbol{\sigma}^{(i)}(\mathbf{x}) + \mathbf{b}^{(i)}(\mathbf{x})=0,
\qquad \mathbf{x}\in\Omega^{(i)}.
\end{equation}
where $\boldsymbol{\sigma}^{(i)}$ is the stress tensor and $\mathbf{b}^{(i)}$ is the prescribed body-force term. The displacement, strain, and stress fields satisfy the small-strain kinematic relation and the linear-elastic constitutive relation,
\begin{equation}\label{eq:constitutive}
\boldsymbol{\varepsilon}^{(i)}
=\frac{1}{2}\left[\nabla\mathbf{u}^{(i)}+
(\nabla\mathbf{u}^{(i)})^{T}\right],
\qquad \boldsymbol{\sigma}^{(i)}=\mathbb{C}:\boldsymbol{\varepsilon}^{(i)}.
\end{equation}
where $\boldsymbol{\varepsilon}^{(i)}$ is the strain tensor and $\mathbb{C}$ is the fourth-order elasticity tensor. All samples within a given dataset share the same material parameters; thus, different load--geometry combinations in that dataset follow the same constitutive relation.

For any $\Omega^{(i)}$, its boundary is partitioned into the Neumann boundary $\Gamma_N^{(i)}$, on which tractions are prescribed, and the Dirichlet boundary $\Gamma_D^{(i)}$, on which displacements are prescribed:
\begin{equation}\label{eq:boundary_partition}
\partial\Omega^{(i)} = \Gamma_N^{(i)}\cup\Gamma_D^{(i)}.
\end{equation}
The corresponding boundary conditions are
\begin{equation}\label{eq:boundary_conditions}
\begin{aligned}
\boldsymbol{\sigma}^{(i)}\mathbf{n}^{(i)} &= \mathbf{f}_N^{(i)}
&&\quad \text{on }\Gamma_N^{(i)},\\
\mathbf{u}^{(i)} &= \mathbf{f}_D^{(i)}
&&\quad \text{on }\Gamma_D^{(i)},
\end{aligned}
\end{equation}
where $\mathbf{n}^{(i)}$ is the outward unit normal vector, $\mathbf{f}_N^{(i)}$ and $\mathbf{f}_D^{(i)}$ denote the Neumann traction function and Dirichlet displacement function, respectively, for the $i$-th sample. The different forms of force excitation and prescribed displacement considered in this study are represented uniformly by the corresponding boundary functions. Geometric variations change the positions, shapes, and normal directions of the internal or external boundaries. Consequently, the loading conditions and geometric configuration jointly define a specific linear-elastic boundary-value problem.

Let $\mathcal{F}_N$ and $\mathcal{F}_D$ denote the admissible function spaces for the Neumann tractions and prescribed Dirichlet displacements, respectively. For a given $(\mathbf{f}_N^{(i)},\mathbf{f}_D^{(i)},\Omega^{(i)})\in\mathcal{F}_N\times\mathcal{F}_D\times\mathfrak{D}$, solving the corresponding linear elastic boundary-value problem, the structural response vector at any query point $\mathbf{x}\in\Omega^{(i)}$ is defined as
\begin{equation}\label{eq:response_vector}
\mathbf{y}^{(i)}(\mathbf{x})=
\begin{bmatrix}
u_x^{(i)}(\mathbf{x}) & u_y^{(i)}(\mathbf{x}) & \sigma_{vm}^{(i)}(\mathbf{x})
\end{bmatrix}^{T}
\end{equation}
where $u_x$ and $u_y$ are the displacement components in the $x$- and $y$-directions, respectively, and $\sigma_{vm}$ is the von Mises stress. Accordingly, the joint load--geometry structural response operator considered in this work can be expressed as
\begin{equation}\label{eq:response_operator}
\mathcal{G}:(\mathbf{f}_N,\mathbf{f}_D,\Omega)\longmapsto\mathbf{y}(\cdot)
\end{equation}
Here, $\mathbf{y}(\cdot)$ denotes the response field defined on $\Omega$. At a query point $\mathbf{x}\in\Omega$, the operator output is evaluated as
\[
\mathcal{G}(\mathbf{f}_N,\mathbf{f}_D,\Omega)(\mathbf{x})=
\begin{bmatrix}
u_x(\mathbf{x}) & u_y(\mathbf{x}) & \sigma_{vm}(\mathbf{x})
\end{bmatrix}^{T}
\]

Our objective is to construct a neural operator $\mathcal{G}_{\Theta}$, parameterized by $\Theta$, that approximates $\mathcal{G}$ over the admissible boundary-function spaces and the family of structural domains:
\begin{equation}\label{eq:approximation}
\mathcal{G}_{\Theta}(\mathbf{f}_N,\mathbf{f}_D,\Omega)(\mathbf{x})
\approx\mathcal{G}(\mathbf{f}_N,\mathbf{f}_D,\Omega)(\mathbf{x}).
\end{equation}
Therefore, LG-DON learns a load--geometry-dependent structural response operator over varying structural domains. This formulation extends beyond a load-to-response mapping on a fixed domain or conventional finite-dimensional regression from geometric parameters to structural responses. Section 2.2 describes the generation of varying physical domains, boundary inputs, and finite element response data. Section 2.3 then presents the LG-DON architecture based on this operator formulation.

\subsection{Dataset Generation and Processing}\label{subsec:dataset}

Finite element simulations were used to generate structural responses for different combinations of geometry and boundary inputs, providing supervised samples of the structural response operator defined in Section 2.1. A geometric parameter vector $\boldsymbol{\gamma}$ was first sampled from the prescribed parameter ranges of each benchmark and used to construct the physical domain $\Omega$. The vector $\boldsymbol{\gamma}$ was used both to generate the finite element geometry and to provide the parameterized geometry input for MIONet.

The positions, sizes, and shapes of irregular holes within the physical domain were varied randomly. The hole centers $(c_x,c_y)$ were generated according to the benchmark-specific rules described in Section 3. Each hole boundary was discretized by $N_h$ points uniformly distributed in polar angle, with $N_h=128$ for the two-hole plate and $N_h=32$ for the cantilever beam. The $j$-th boundary point was defined as
\begin{equation}\label{eq:hole_boundary}
\begin{aligned}
x_j &= c_x+r(\theta_j)\cos\theta_j,\\
y_j &= c_y+r(\theta_j)\sin\theta_j,
\end{aligned}
\end{equation}
where $\theta_j=\frac{2\pi j}{N_h}$, $j=0,\ldots,N_h-1$. The radial function $r(\theta_j)$ was constructed from random Fourier modes of orders two through seven, with a random reference radius and perturbation amplitude controlling the hole size and shape. For the L-shaped bracket with a varying external geometry, different physical domains were constructed by varying the geometric parameters associated with the outer boundary. The geometric dimensions, parameter ranges, and constraints for each benchmark are specified in Section 3.

The boundary inputs considered in this study comprised distributed traction, concentrated forces, and prescribed displacements applied either over an entire designated boundary or over a localized boundary segment. These inputs were represented by the Neumann traction function $\mathbf{f}_N^{(i)}$ and the Dirichlet displacement function $\mathbf{f}_D^{(i)}$ defined in Section 2.1. Gaussian random fields (GRFs) were used to generate distributed traction and full-boundary prescribed displacement, producing continuously varying input functions along the designated boundaries. Concentrated forces were represented using finite-order Fej\'er kernels as smooth approximations to ideal Dirac-type loads. Localized prescribed displacement was constructed over a finite support interval using a Fej\'er-weighted Fourier series. The input amplitudes, application locations, and other control parameters were sampled randomly from the prescribed ranges of each benchmark.

Each triplet $(\Omega^{(i)},\mathbf{f}_N^{(i)},\mathbf{f}_D^{(i)})$ defined an independent static linear-elastic boundary-value problem. An unstructured triangular mesh was generated directly on the corresponding physical domain $\Omega^{(i)}$, and the problem was solved using the finite element method (FEM). Because the geometries differed between samples, both the nodal coordinates and the total number of nodes $N_i$ could vary. The coordinates of finite element node $n$ in sample $i$ were denoted by
\begin{equation}\label{eq:fem_nodes}
\mathbf{x}_n^{(i)}=(x_n^{(i)},y_n^{(i)}),\qquad n=1,\ldots,N_i.
\end{equation}
Following Eq. (7) in Section 2.1, the corresponding supervised response was written as
\[
\mathbf{y}_n^{(i)}=
\begin{bmatrix}
u_{x,n}^{(i)} & u_{y,n}^{(i)} & \sigma_{vm,n}^{(i)}
\end{bmatrix}^{T}
\]
The raw data retained for finite element sample $i$ were therefore represented as
\begin{equation}\label{eq:raw_sample}
\mathcal{S}^{(i)}=\left\{\boldsymbol{\gamma}^{(i)},\Omega^{(i)},\mathbf{f}_N^{(i)},\mathbf{f}_D^{(i)},(\mathbf{x}_n^{(i)},\mathbf{y}_n^{(i)})_{n=1}^{N_i}\right\}
\end{equation}
All nodal coordinates and response fields were retained on the original unstructured finite element mesh without additional interpolation onto a common Cartesian grid. This avoided additional loss of geometric and response information caused by interpolation near internal holes and varying boundaries. After the supervised data had been generated, the boundary functions were sampled at sensor locations to construct the neural operator inputs. Geometry inputs were then prepared according to the different geometry representations used by modified MIONet, introduced in Section 2.3.1, and LG-DON.

The continuous Neumann and Dirichlet input functions on the designated boundaries were sampled at $N_s$ sensor locations $\{s_l\}_{l=1}^{N_s}$. The discrete boundary inputs for sample $i$ were defined as
\begin{equation}\label{eq:discrete_boundary_inputs}
\begin{aligned}
F_N^{(i)}&=\left[\mathbf{f}_N^{(i)}(s_1),\mathbf{f}_N^{(i)}(s_2),\ldots,\mathbf{f}_N^{(i)}(s_{N_s})\right],\\
F_D^{(i)}&=\left[\mathbf{f}_D^{(i)}(s_1),\mathbf{f}_D^{(i)}(s_2),\ldots,\mathbf{f}_D^{(i)}(s_{N_s})\right],
\end{aligned}
\end{equation}
In this study, $N_s=101$ boundary sensors were used. The finite element nodes $\mathbf{x}_n^{(i)}$ served directly as query points, with the corresponding responses $\mathbf{y}_n^{(i)}$ used as supervised targets.

Before training, the variables were normalized to reduce differences in numerical scale between physical quantities:
\begin{equation}\label{eq:normalization}
\bar q=\frac{q}{\max\lvert q\rvert}
\end{equation}
Here, $q$ denotes a variable such as a spatial coordinate, a boundary input, or a structural response component, and $\bar q$ is its normalized, dimensionless counterpart. Each physical component was normalized using its corresponding reference scale. For a given variable, this scale was kept identical during training, validation, and testing, and remained fixed in the cross-discretization and extrapolation experiments. The network outputs were subsequently transformed back to dimensional structural responses by inverse normalization.

\subsection{Model Architectures}\label{subsec:architectures}

Following the preprocessing described in Section 2.2, all network inputs are dimensionless unless otherwise stated. To address structural response prediction under varying loads and geometries, we first introduce the multiple-input operator network (MIONet) and then describe the proposed global--local geometry representation in Load- and Geometry-adaptive Deep Operator Network (LG-DON). All models in this study were trained using normalized quantities, and their predicted responses were restored to physical units by inverse normalization.

\subsubsection{MIONet Architecture}\label{subsubsec:mionet}

\begin{figure}[pos=!htbp]
\centering
\includegraphics[width=\linewidth]{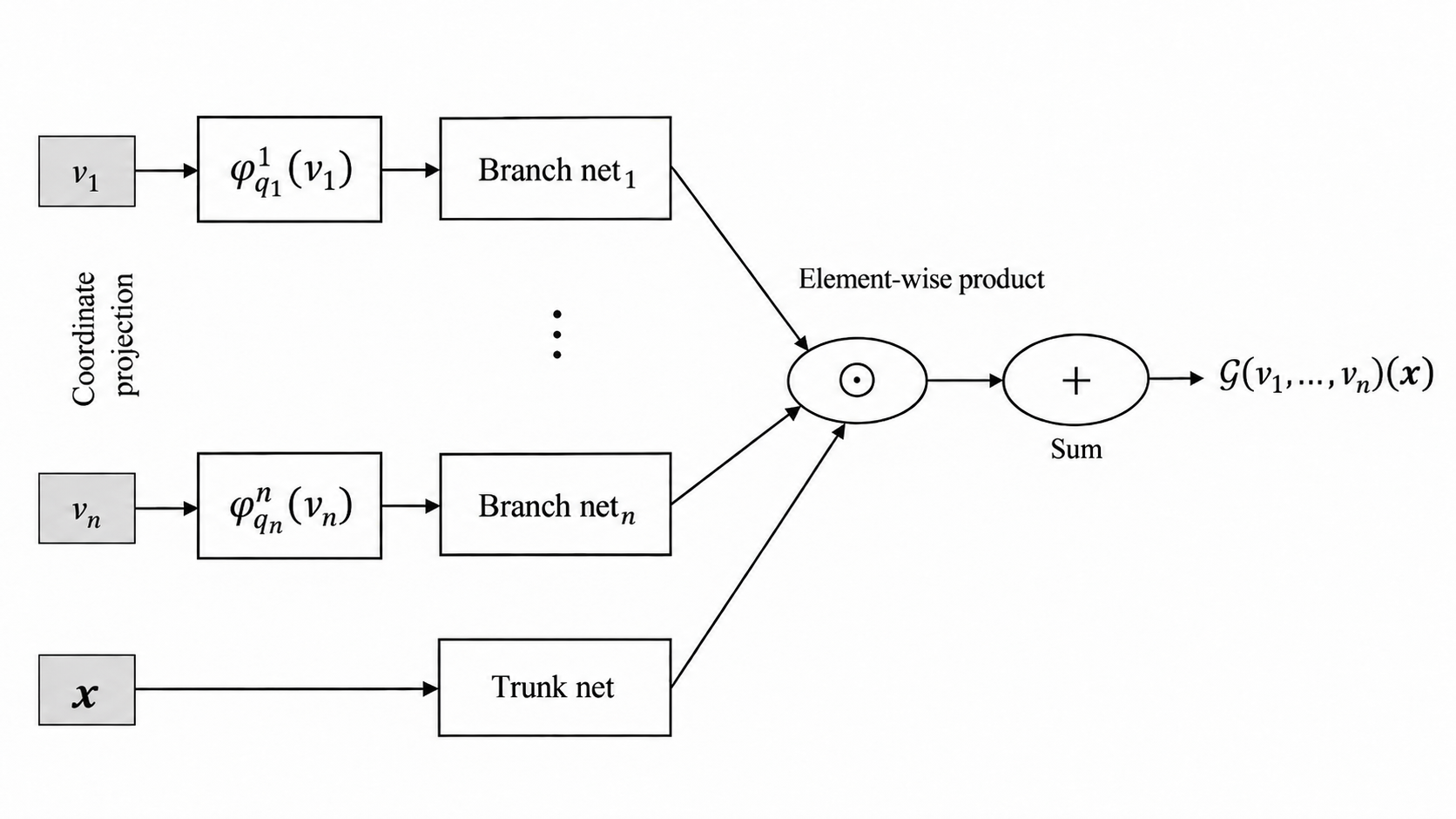}
\caption{MIONet architecture}
\label{fig:mionet}
\end{figure}

MIONet extends Deep Operator Network (DeepONet) to problems in which multiple independent input functions jointly determine an output function. A standard low-rank MIONet consists of multiple branch networks and a trunk network. Each branch network encodes one input function, while the trunk network encodes the query coordinates of the output function. As illustrated in Fig.~\ref{fig:mionet}, the branch and trunk networks produce latent feature vectors of equal dimension. These vectors are fused elementwise using the Hadamard product and then summed over the latent dimension to obtain a scalar prediction at the query point.

\begin{equation}
\label{eq:mionet_standard}
\mathcal{G}_{\Theta}(v_1,v_2,\cdots,v_m)(\mathbf{x})
=
S\left(
\mathcal{B}_1(v_1)\odot\cdots\odot
\mathcal{B}_m(v_m)\odot T(\mathbf{x})
\right)+b_0
\end{equation}
Here, $\odot$ denotes the Hadamard product, and $S$ denotes summation over all vector components.

\begin{figure}[pos=!htbp]
\centering
\includegraphics[width=\linewidth]{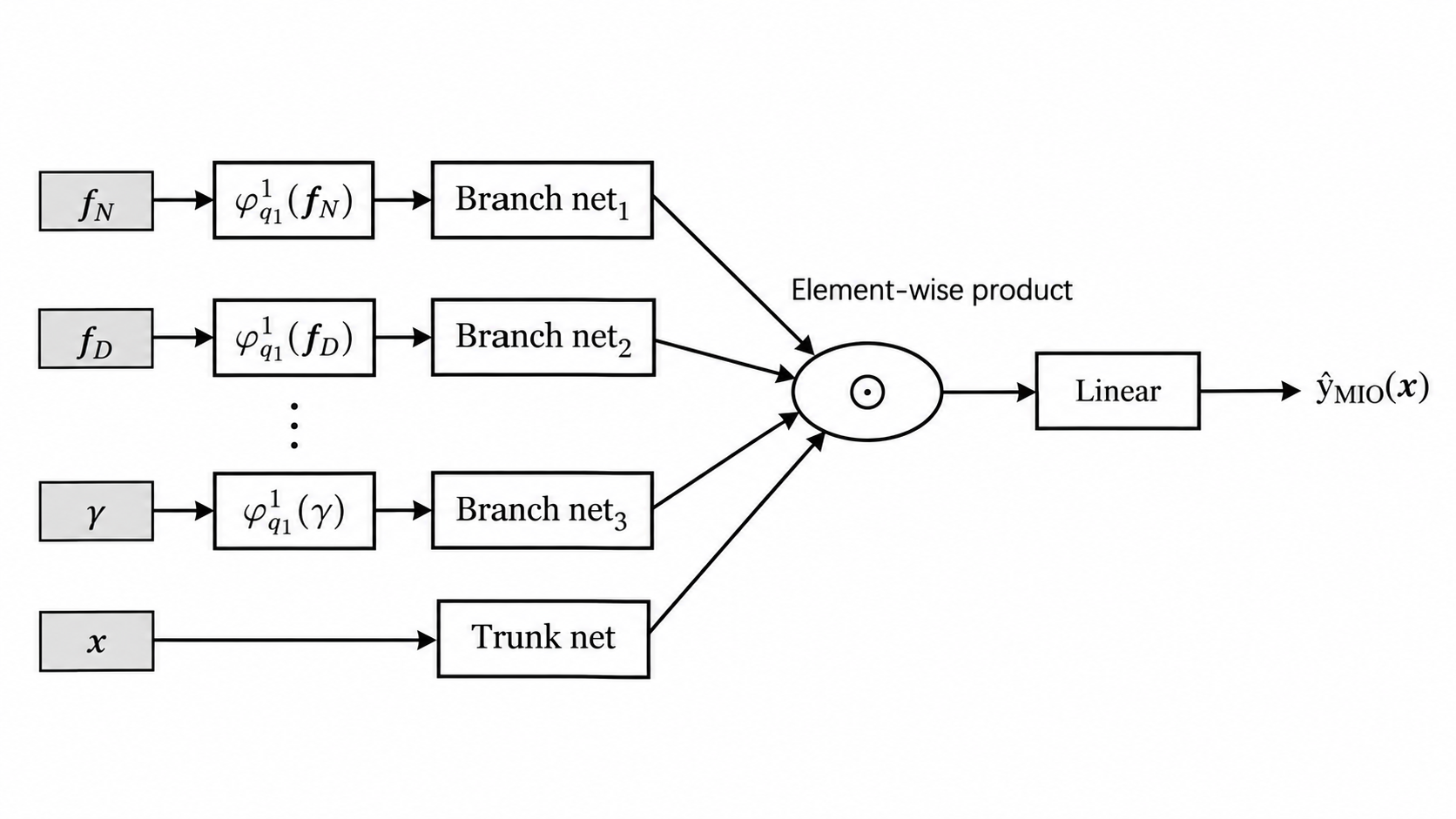}
\caption{Modified MIONet architecture.}
\label{fig:modified_mionet}
\end{figure}

To predict three structural response components, we modify the low-rank MIONet architecture. As shown in Fig.~\ref{fig:modified_mionet}, modified MIONet retains the latent feature dimension after fusing the branch and trunk features through the Hadamard product. A single fully connected output layer then predicts the two displacement components and the von Mises stress:

\begin{equation}
\label{eq:modified_mionet}
\begin{aligned}
\mathbf{h}_{\mathrm{MIO}}
&=
\left(
\mathcal{B}_1(v_1)\odot\cdots\odot
\mathcal{B}_m(v_m)\odot T(\mathbf{x})
\right)+b_0,\\
\widehat{\mathbf{y}}_{\mathrm{MIO}}
&=
\mathbf{W}_{out}\mathbf{h}_{\mathrm{MIO}}
+\mathbf{b}_{out},
\quad
\mathbf{W}_{out}\in\mathbb{R}^{3\times q},
\quad
\mathbf{b}_{out}\in\mathbb{R}^{3},
\end{aligned}
\end{equation}
here, $\widehat{\mathbf{y}}_{\mathrm{MIO}}=
\begin{bmatrix}
\widehat{u}_x &
\widehat{u}_y &
\widehat{\sigma}_{vm}
\end{bmatrix}^{T}$,and the geometry branch of modified MIONet takes the auxiliary parameter vector $\boldsymbol{\gamma}$ used to construct the physical domain in Section 2.2. Its mapping for structural response prediction under varying loads and geometries is therefore expressed as

\begin{equation}
\label{eq:mionet_operator}
\begin{aligned}
&\mathbf{h}_{\mathrm{MIO}}
=
\left(
\mathcal{B}_1(\mathbf{f}_N)
\odot
\mathcal{B}_2(\mathbf{f}_D)
\odot
\mathcal{B}_3(\boldsymbol{\gamma})
\odot
T(\mathbf{x})
\right)+b_0,\\
&\mathcal{G}_{\Theta}^{\mathrm{MIO}}
(\mathbf{f}_N,\mathbf{f}_D,\boldsymbol{\gamma})(\mathbf{x})
=
\mathbf{W}_{out}\mathbf{h}_{\mathrm{MIO}}
+\mathbf{b}_{out},
\quad
\mathbf{W}_{out}\in\mathbb{R}^{3\times q},
\quad
\mathbf{b}_{out}\in\mathbb{R}^{3},\\
&\widehat{\mathbf{y}}_{\mathrm{MIO}}(\mathbf{x})
=
\mathcal{G}_{\Theta}^{\mathrm{MIO}}
(F_N,F_D,\boldsymbol{\gamma})(\mathbf{x}).
\end{aligned}
\end{equation}

\subsubsection{Joint DG--SDF geometry encoding}\label{subsubsec:dg_sdf}

Whereas modified MIONet uses a geometric parameter vector, LG-DON constructs its geometry representation directly from discrete geometry point sets and query-point-to-boundary distances. For the varying computational domain $\Omega^{(i)}$ of sample $i$, the discrete geometry encoder (DG) extracts sample-level global geometry features that are independent of the query location. The SDF-based boundary-distance features describe the spatial relationships between individual query points and selected physical boundaries. Together, these features form the global--local geometry representation of LG-DON.

\noindent\textbf{(1) Discrete geometry encoding}

For the discrete geometry points $\mathbf{x}^{(i)}$ of sample $i$, three candidate DG point sets are defined according to the geometric information they contain:

\begin{equation}
\label{eq:dg_point_sets}
\begin{array}{l@{\quad}c@{\quad}l}
\mathcal{P}_{vary}^{(i)}
& = &
\left\{
\mathbf{x}^{(i)},\,
\mathbf{x}^{(i)}\in\Gamma_{vary}^{(i)}
\right\},\\
\mathcal{P}_{bound}^{(i)}
& = &
\left\{
\mathbf{x}^{(i)},\,
\mathbf{x}^{(i)}\in\partial\Omega^{(i)}
\right\},\\
\mathcal{P}_{domain}^{(i)}
& = &
\left\{
\mathbf{x}^{(i)},\,
\mathbf{x}^{(i)}\in\Omega^{(i)}
\right\},
\end{array}
\end{equation}
the three sets comprise points on the varying boundaries $\Gamma_{vary}^{(i)}$, points on all boundaries $\partial\Omega^{(i)}$, and all discrete points in the domain $\Omega^{(i)}$, respectively. For a selected configuration $c\in\{\mathrm{vary\_bound},\mathrm{all\_bound},\mathrm{all\_domain}\}$, the corresponding point set is denoted by

\[
\mathcal{P}_{c}^{(i)}
=
\left\{
\mathbf{x}_{j}^{(i)}:
j=1,\ldots,m
\right\}
\]
here, $m$ is the number of selected geometry points and may vary between samples. After normalization as described in Section 2.2, the point coordinates are transformed by a shared pointwise mapping. A pooling layer then aggregates the pointwise features into a fixed-dimensional geometry representation:

\begin{equation}
\label{eq:dg_global_feature}
\mathbf{g}^{(i)}
=
\mathrm{pool}
\left(
\left\{
H(\mathcal{P}_{c}^{(i)})
\right\}
\right)
\end{equation}
here, $H$ is a trainable shared pointwise mapping. The dimension of $\mathbf{g}^{(i)}$ is independent of the number of geometry points, and the representation is invariant to their ordering. The resulting feature $\mathbf{g}^{(i)}$ is independent of the query location and is shared by all query points in the same sample. It represents hole positions and contours, the outer boundary shape, and the relative arrangement of geometric components. The effects of the three point-selection strategies are compared through ablation studies in Section 4.

\noindent\textbf{(2) SDF-based boundary-distance feature}

To complement the sample-level DG representation with query-dependent geometry information, we construct distances from each query point to predefined physical boundary categories. For sample $i$, let $\Gamma_{k}^{(i)}$ denote the $k$-th boundary category, with

\[
\partial\Omega^{(i)}
=
\bigcup_{k=1}^{K}\Gamma_{k}^{(i)},
\]
where $K$ is the number of boundary categories specified for the geometry family. The minimum Euclidean distance from query point $\mathbf{x}_{n}^{(i)}$ to category $\Gamma_{k}^{(i)}$ is

\begin{equation}
\label{eq:boundary_distance}
d_{k}^{(i)}(\mathbf{x}_{n}^{(i)})
=
\min_{\boldsymbol{\xi}\in\Gamma_{k}^{(i)}}
\left\|
\mathbf{x}_{n}^{(i)}-\boldsymbol{\xi}
\right\|_2,
\qquad k=1,\ldots,K
\end{equation}
this distance is nonnegative and vanishes when the query point lies on the corresponding boundary category.

A conventional signed distance function represents the shortest distance to a boundary together with a sign distinguishing its two sides. Here, all query points lie within the material domain or on its boundary, and only nonnegative distance magnitudes are used. We therefore refer to these inputs as SDF-based boundary-distance features rather than a signed distance field defined over the ambient space.

Building on the scalar SDF conditioning discussed for Geom-DeepONet \cite{he2024lgdon20}, we compare two distance representations, denoted by outer and boundary:

\begin{equation}
\label{eq:sdf_configurations}
d_{s}^{(i)}(\mathbf{x})
=
\begin{cases}
\left[
\displaystyle\min_{k=1,\ldots,K}
d_{k}^{(i)}(\mathbf{x})
\right],
& s=\mathrm{outer},\\[6pt]
\left[
d_{1}^{(i)}(\mathbf{x}),\ldots,
d_{K}^{(i)}(\mathbf{x})
\right],
& s=\mathrm{boundary}.
\end{cases}
\end{equation}

Both configurations cover the complete domain boundary, including internal hole boundaries where present. The outer configuration retains a single nearest-boundary distance; the label outer does not imply that hole boundaries are excluded. The boundary configuration retains the distance to each predefined category through a fixed channel assignment. It therefore preserves both the association between channels and boundary categories and the distances to categories other than the nearest one.

For configuration $s$, the distance features are concatenated with the query coordinates to form the query-level geometry input:

\begin{equation}
\label{eq:query_geometry_input}
z_{n,s}^{(i)}
=
x_{n}^{(i)}
\mathbin{\|}
d_{s}^{(i)}(x_{n}^{(i)}),
\end{equation}
where $\|$ denotes concatenation. The number of distance channels is $K_s=1$ for outer and $K_s=K$ for boundary. After normalization, this input is supplied to the geometry-aware trunk pathway. Together, $\mathbf{g}^{(i)}$ and $z_{n,s}^{(i)}$ provide sample-level geometry information and query-dependent boundary relationships, respectively.

\subsubsection{LG-DON Architecture}\label{subsubsec:lgdon}

\begin{figure}[pos=!htbp]
\centering
\includegraphics[width=\linewidth]{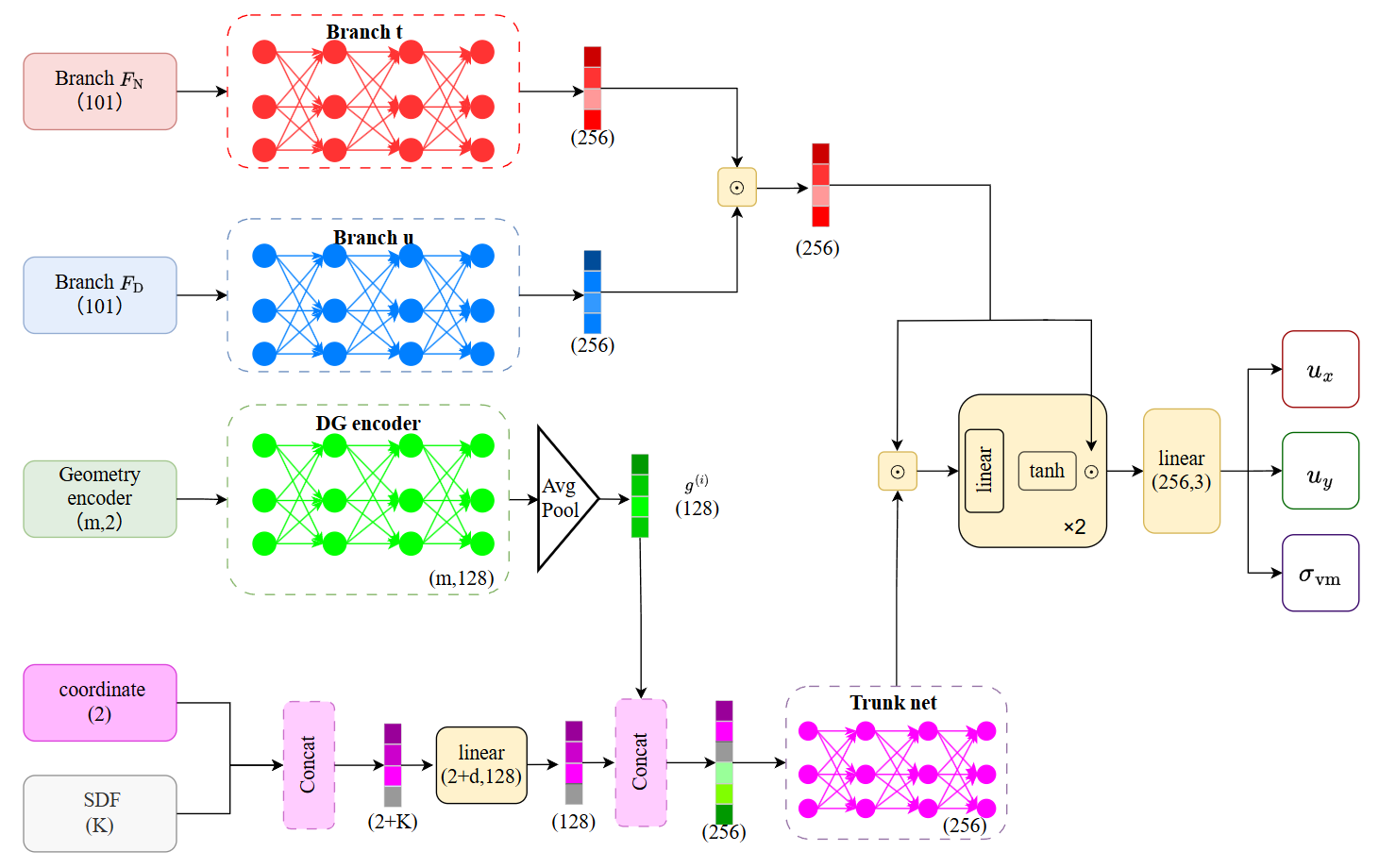}
\caption{The architecture of the LG-DON.}
\label{fig:lgdon}
\end{figure}

The overall LG-DON architecture is shown in Fig.~\ref{fig:lgdon}. The model takes the discrete Neumann traction data $\mathbf{f}_N$ and Dirichlet displacement data $\mathbf{f}_D$ defined in Section 2.2, together with the DG point set and the combined query-coordinate and boundary-distance input constructed in Section 2.3.2. The predicted structural response is expressed as

\begin{equation}
\label{eq:lgdon_operator}
\widehat{\mathbf{y}}_{\mathrm{LG}}(\mathbf{x})
=
\mathcal{G}_{\Theta}^{\mathrm{LG}}
(\mathbf{f}_N,\mathbf{f}_D,\mathcal{P}_c)(\mathbf{x})
\end{equation}
here, $\Theta$ denotes the trainable network parameters, and $c$ identifies the DG point-selection strategy.

The two boundary inputs are encoded by separate branch networks. Their representations are fused through the Hadamard product to form a joint load feature

\begin{equation}
\label{eq:joint_load_feature}
\mathbf{e}^{(i)}
=
\mathcal{B}_N(\mathbf{f}_N^{(i)})
\odot
\mathcal{B}_D(\mathbf{f}_D^{(i)}).
\end{equation}
here, $\mathcal{B}_N(\mathbf{f}_N^{(i)}),
\mathcal{B}_D(\mathbf{f}_D^{(i)})\in\mathbb{R}^{q}$, and hence $\mathbf{e}^{(i)}\in\mathbb{R}^{q}$. This feature is shared by all query points in the sample. For samples with only one type of boundary excitation, the inactive branch receives an all-zero input vector. This zero vector is an input-encoding convention and must be distinguished from a homogeneous Dirichlet condition imposed in the FEM model. The actual boundary constraints follow the scenario-specific definitions in Section 3. Both boundary inputs are retained for samples subjected to combined loading.

Along the geometry pathway, the DG encoder produces the sample-level global geometry feature $\mathbf{g}^{(i)}$ described in Section 2.3.2. The combined query-coordinate and boundary-distance input is transformed by a linear lifting layer and a hyperbolic tangent activation into a query-level feature:

\begin{equation}
\label{eq:query_lifting}
\boldsymbol{\ell}_{n}^{(i)}
=
\tanh
\left(
\mathbf{W}_{lift}z_{n}^{(i)}
+\mathbf{b}_{lift}
\right)
\end{equation}

The sample-level global geometry feature and the query-level feature are then concatenated to form the geometry-aware input to the trunk network

\begin{equation}
\label{eq:trunk_geometry_input}
\mathbf{a}_{n}^{(i)}
=
\mathbf{g}^{(i)}
\mathbin{\|}
\boldsymbol{\ell}_{n}^{(i)}
\end{equation}

It is worth noting that the two branch networks, labeled Branch $F_N$ and Branch $F_D$, and the first stage of the trunk network use modified multilayer perceptrons (MMLPs) \cite{wang2020lgdon34} for feature transformation.

\begin{equation}
\label{eq:mmlp_uv}
\begin{aligned}
\mathbf{U}
&=
\Phi(\mathbf{X}\mathbf{W}^{1}+\mathbf{b}^{1}),\\
\mathbf{V}
&=
\Phi(\mathbf{X}\mathbf{W}^{2}+\mathbf{b}^{2})
\end{aligned}
\end{equation}

\begin{equation}
\label{eq:mmlp_h1}
\mathbf{H}^{1}
=
\Phi
\left(
\mathbf{X}\mathbf{W}^{z,1}
+\mathbf{b}^{z,1}
\right)
\end{equation}

\begin{equation}
\label{eq:mmlp_z}
\mathbf{Z}^{(l)}
=
\Phi
\left(
\mathbf{H}^{l}\mathbf{W}^{z,l}
+\mathbf{b}^{z,l}
\right),
\qquad l=1,\cdots,L
\end{equation}

\begin{equation}
\label{eq:mmlp_h}
\mathbf{H}^{(l+1)}
=
(1-\mathbf{Z}^{(l)})\odot\mathbf{U}
+
\mathbf{Z}^{(l)}\odot\mathbf{V},
\qquad l=1,\cdots,L
\end{equation}

\begin{equation}
\label{eq:mmlp_output}
f_{\Theta}(x)
=
\mathbf{H}^{(L+1)}\mathbf{W}
+\mathbf{b}
\end{equation}

The joint load feature is broadcast across the query-point dimension and progressively fused with the geometry-aware features through Hadamard products at multiple stages of the trunk network.

Finally, a single shared trunk network and a shared linear output layer jointly predict the two displacement components and the von Mises stress:

\begin{equation}
\label{eq:lgdon_output}
\widehat{\mathbf{y}}_{\mathrm{LG},n}^{(i)}
=
\begin{bmatrix}
u_{x,n}^{(i)} &
u_{y,n}^{(i)} &
\sigma_{vm,n}^{(i)}
\end{bmatrix}^{T}
\end{equation}

Compared with modified MIONet in Section 2.3.1, LG-DON replaces the geometric-parameter branch with DG-derived global geometry features and query-dependent boundary-distance features. It also incorporates the joint load feature at multiple stages of the trunk network, enabling prediction of multiple structural response fields under simultaneous variations in boundary inputs and geometry.
\FloatBarrier

\section{Numerical Experiments and Results}\label{sec:experiments}

We evaluated LG-DON for structural response prediction using the three two-dimensional elasticity benchmarks with varying geometries summarized in table.~\ref{tab:numerical_cases}. Case 1 considers a plate with two irregular holes and multiple types of boundary input. This case assesses the adaptability of LG-DON, examines the effects of discretization on neural operator performance, and provides a comparison with modified MIONet. Case 2 considers a cantilever beam with a single irregular hole and evaluates both in-distribution prediction and extrapolation in load amplitude and geometric parameters. Case 3 considers an L-shaped bracket with a varying external geometry and assesses prediction performance as the material domain and boundary locations change. The first two cases focus on variations in internal boundaries within a fixed outer domain, whereas the third extends the study to varying outer boundaries.

All physical quantities are expressed in SI units: lengths in m, forces in N, and stresses in Pa. All finite element (FEM) data were generated using MATLAB on a computer with 16 GB of RAM and an Intel Core i9-13980HX CPU operating at 2.20 GHz. LG-DON was implemented in PyTorch and trained on an NVIDIA RTX 4060 GPU. Both LG-DON and modified MIONet have approximately 1.1 million trainable parameters, with slight variations across geometry-encoding configurations. To ensure a fair comparison, the same training settings were used for all cases and models. The Adam optimizer was used with an initial learning rate of $10^{-4}$. The learning rate was multiplied by 0.5 whenever the validation metric failed to improve for 50 consecutive epochs. The hyperbolic tangent (tanh) activation function was used. For each case, the dataset was divided into training, validation, and test sets in a 7:1:2 ratio.

\begin{table}[pos=!htbp]
\centering
\caption{Overview of the numerical cases}
\label{tab:numerical_cases}

\footnotesize
\setlength{\tabcolsep}{3pt}
\renewcommand{\arraystretch}{1.15}

\begin{tabularx}{\linewidth}{@{}>{\raggedright\arraybackslash}p{0.16\linewidth}>{\raggedright\arraybackslash}p{0.14\linewidth}>{\raggedright\arraybackslash}X>{\centering\arraybackslash}p{0.065\linewidth}>{\raggedright\arraybackslash}p{0.23\linewidth}@{}}
\toprule
Case & Geometric variation & Loading conditions & Samples & Main purpose \\
\midrule
Irregular-hole plate
&
Two varying internal holes
&
Distributed traction, concentrated force, global/local imposed displacement
&
2000
&
Multiple excitation types, output-discretization study, and baseline comparison
\\

Cantilever beam
&
One varying internal hole
&
Distributed traction, imposed displacement, combined loading
&
1500
&
Interpolation and load/geometry extrapolation
\\

L-shaped bracket
&
Varying external boundary
&
Distributed traction and localized imposed displacement
&
1000
&
Varying external geometry
\\
\bottomrule
\end{tabularx}
\end{table}

During training, the discrepancy between the predicted responses and the FEM reference data was quantified using the mean squared error (MSE) loss:

\begin{equation}
\label{eq:mse_loss}
\mathcal{L}_{\mathrm{MSE}}
=
\frac{1}{B}
\sum_{i=1}^{B}
\frac{1}{N_i}
\sum_{j=1}^{N_i}
\left\|
\widehat{\mathbf{y}}_{ij}
-
\mathbf{y}_{ij}
\right\|_2^2
\end{equation}
here, $B$ is the number of samples in a mini-batch, $N_i$ is the number of valid nodes in geometry sample $i$, and $\widehat{\mathbf{y}}$ and $\mathbf{y}$ denote the normalized network predictions and FEM responses, respectively. During validation and testing, performance was assessed using samplewise relative $L_2$ errors. The mean relative error for response field $c$ is defined as:

\begin{equation}
\label{eq:relative_l2_error}
E_{\alpha}
=
\frac{1}{N_{test}}
\sum_{i=1}^{N_{test}}
\frac{
\left\|
\widehat{\mathbf{y}}_{\alpha}^{(i)}
-
\mathbf{y}_{\alpha}^{(i)}
\right\|_2
}{
\left\|
\mathbf{y}_{\alpha}^{(i)}
\right\|_2
}
\end{equation}

The overall evaluation metric is the arithmetic mean of the mean relative $L_2$ errors for the three response fields:

\begin{equation}
\label{eq:mean_relative_error}
E_{mean}
=
\frac{1}{3}
\left(
E_{u_x}
+
E_{u_y}
+
E_{\sigma_{vm}}
\right)
\end{equation}

Unless otherwise stated, Section 3 reports results from a single training run per model and setting, with LG-DON using the \texttt{all\_bound+\allowbreak{}boundary} configuration selected in Section~\ref{sec:ablation}. Section~\ref{sec:ablation} instead reports averages over three independent runs for each geometry-encoding configuration.

\subsection{Elastic Plate with two Irregular Holes}\label{subsec:plate}

We consider a two-dimensional isotropic square plate with a side length of 1 m and two irregular holes to evaluate LG-DON under variations in boundary input type, internal geometry, and discretization. A plane-stress assumption was adopted, with Young's modulus $E = 10$ Pa and Poisson's ratio $v = 0.25$. The bottom edge was fully fixed, while the left and right outer edges and both hole boundaries were traction-free. Boundary inputs were applied to the top edge. The irregular holes were generated using Eq.~\eqref{eq:hole_boundary} in Section 2.2. The radial distance was restricted to $[0.075, 0.12]$ m. The nominal hole centers were $(-0.25, 0)$ and $(0.25, 0)$, with small random perturbations in both the $x$- and $y$-directions. For modified MIONet, the hole geometries were represented by a finite-dimensional parameter vector comprising the center coordinates, radial scale parameters, and Fourier shape coefficients of both holes. The parameters of the two holes were concatenated to form a sample-level geometry input. In contrast, LG-DON does not rely on predefined geometric parameters. It uses the discrete geometry encoder (DG) to extract global geometry features and combines these with query-dependent SDF-based boundary-distance features.

\begin{figure}[pos=!htbp]
\centering
\includegraphics[width=\linewidth]{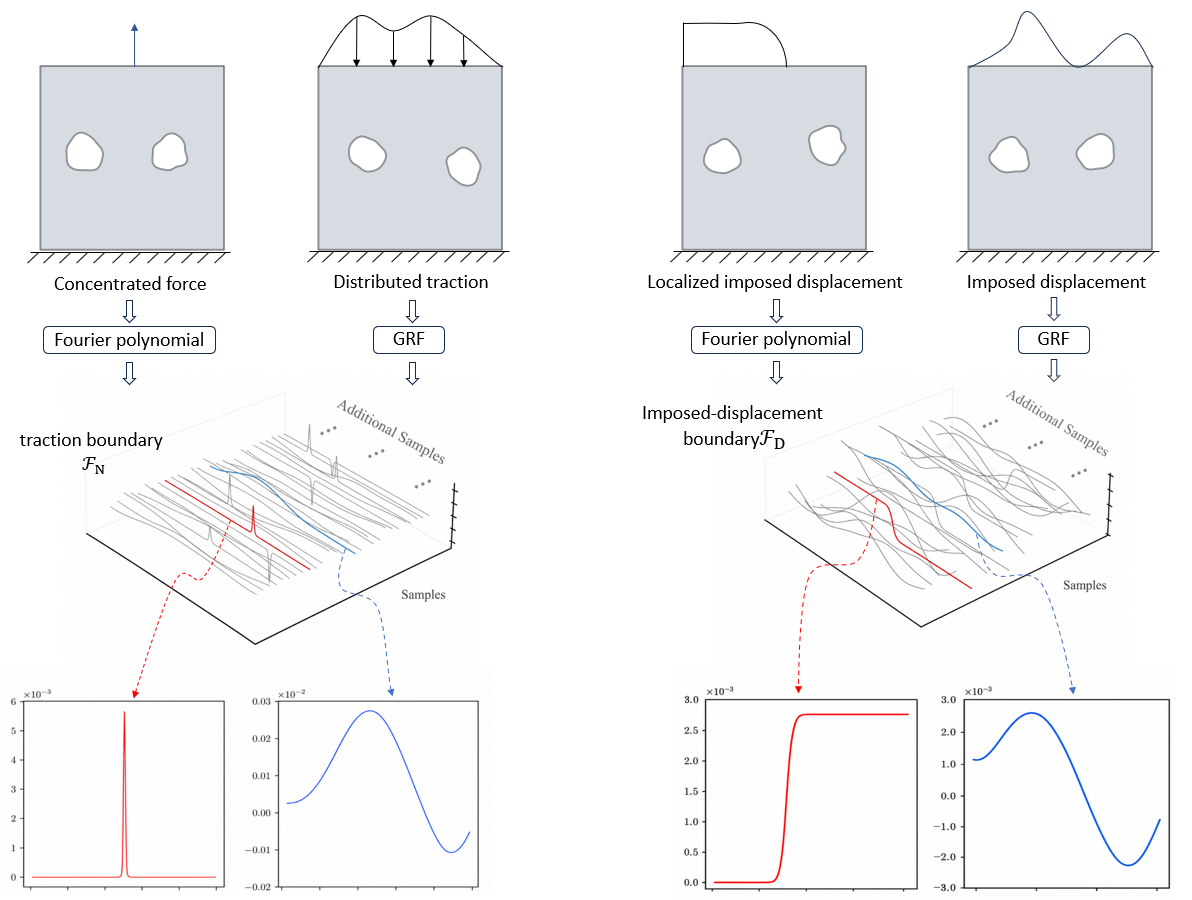}
\caption{Schematic of force and prescribed-displacement input functions generated using GRFs and Fourier-based methods.}
\label{fig:boundary_inputs}
\end{figure}

As illustrated in Fig.~\ref{fig:boundary_inputs}, four types of boundary input were considered: concentrated forces, distributed traction, full-boundary prescribed displacement, and localized prescribed displacement. Each boundary input function was sampled at 101 equally spaced sensor locations.

\begin{figure}[pos=!htbp]
\centering
\includegraphics[width=0.95\linewidth]{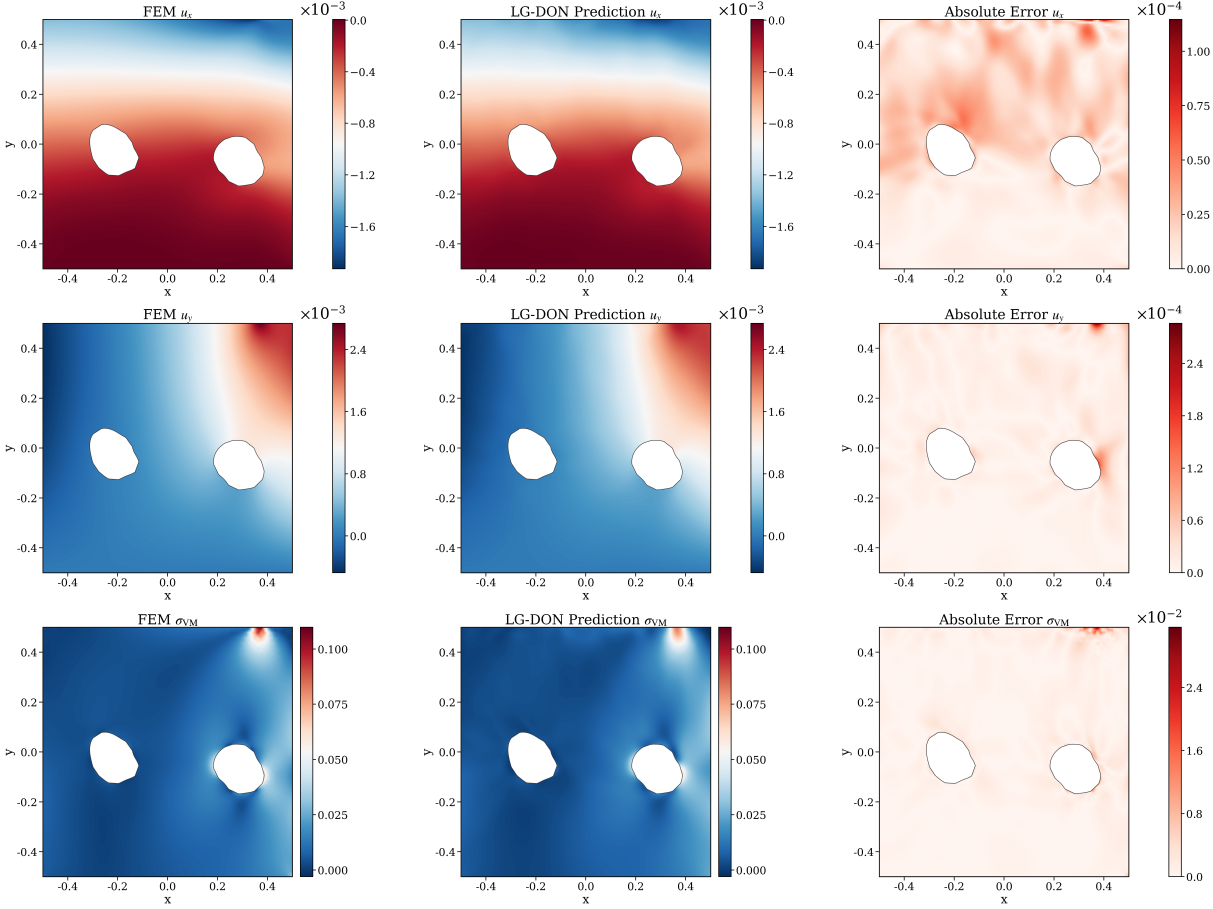}
\caption{Predictions under concentrated-force loading.}
\label{fig:concentrated_force}
\end{figure}

\begin{figure}[pos=!htbp]
\centering
\includegraphics[width=0.95\linewidth]{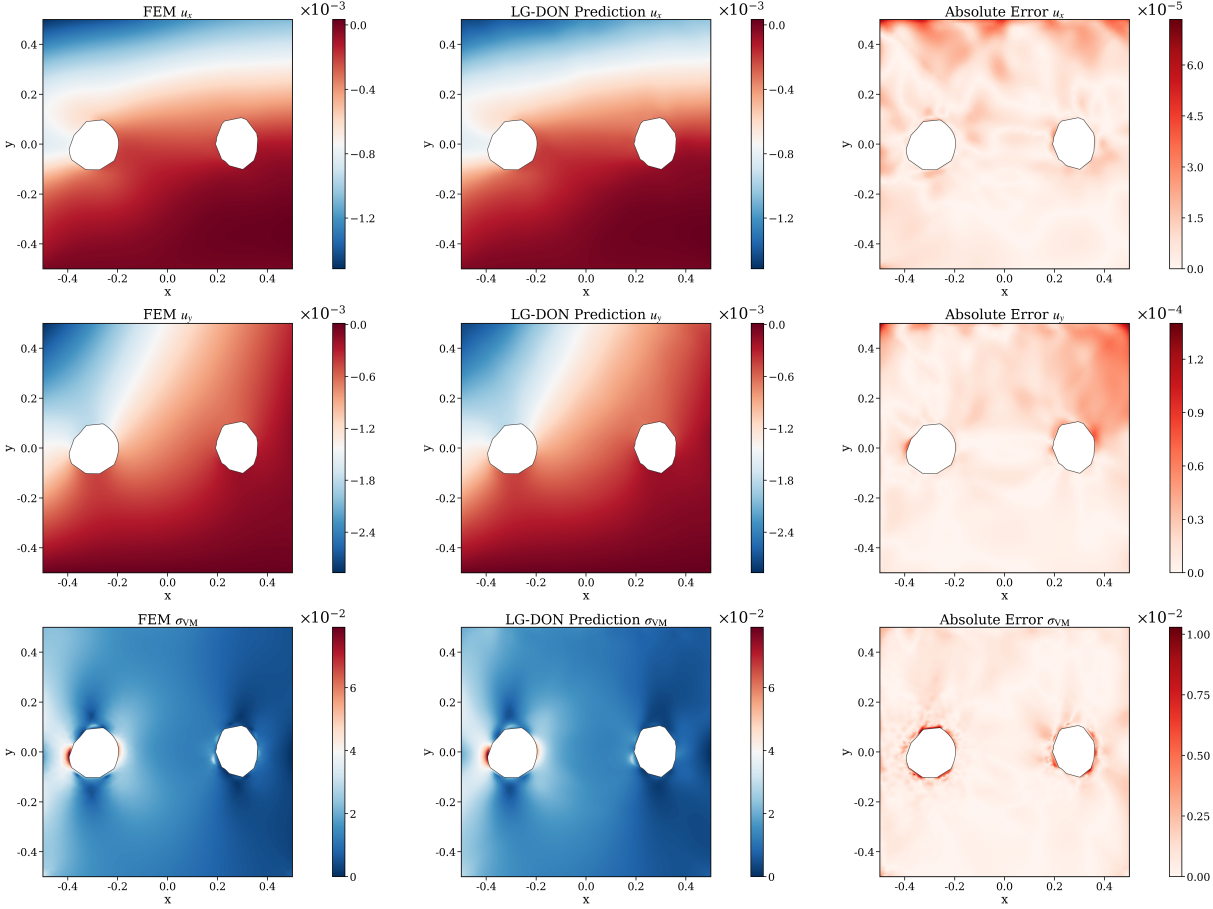}
\caption{Predictions under distributed-traction loading.}
\label{fig:distributed_traction}
\end{figure}

\begin{figure}[pos=!htbp]
\centering
\includegraphics[width=0.95\linewidth]{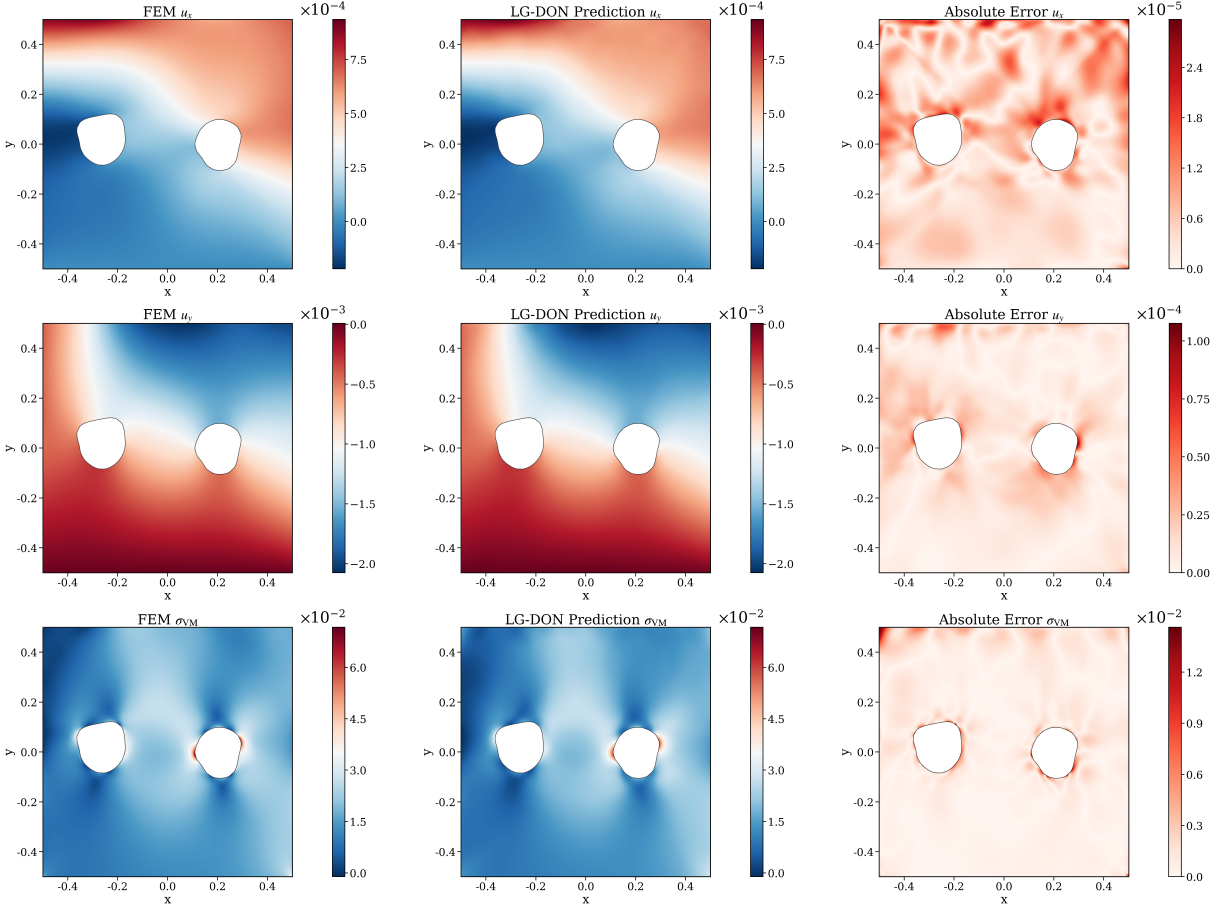}
\caption{Prediction results under global imposed-displacement loading.}
\label{fig:global_displacement}
\end{figure}

After training, the network parameters were held fixed, and predictions were obtained for all four loading conditions without retraining or fine-tuning. Fig.~\ref{fig:concentrated_force} and Fig.~\ref{fig:distributed_traction} show representative results for concentrated-force and distributed-traction loading, respectively. LG-DON reproduced the overall spatial distributions of $u_x$, $u_y$, and von Mises stress. Compared with smooth distributed traction, a concentrated force produces stronger local response gradients near its application point. Local field variations around the hole boundaries further increase the spatial complexity of the response. The prediction errors were predominantly located near the load application region and the hole boundaries. Fig.~\ref{fig:global_displacement} further shows that, under nonhomogeneous Dirichlet displacement boundary conditions, the model reproduced the overall displacement field and the stress concentrations near the holes

\begin{figure}[pos=!htbp]
\centering
\includegraphics[width=\linewidth]{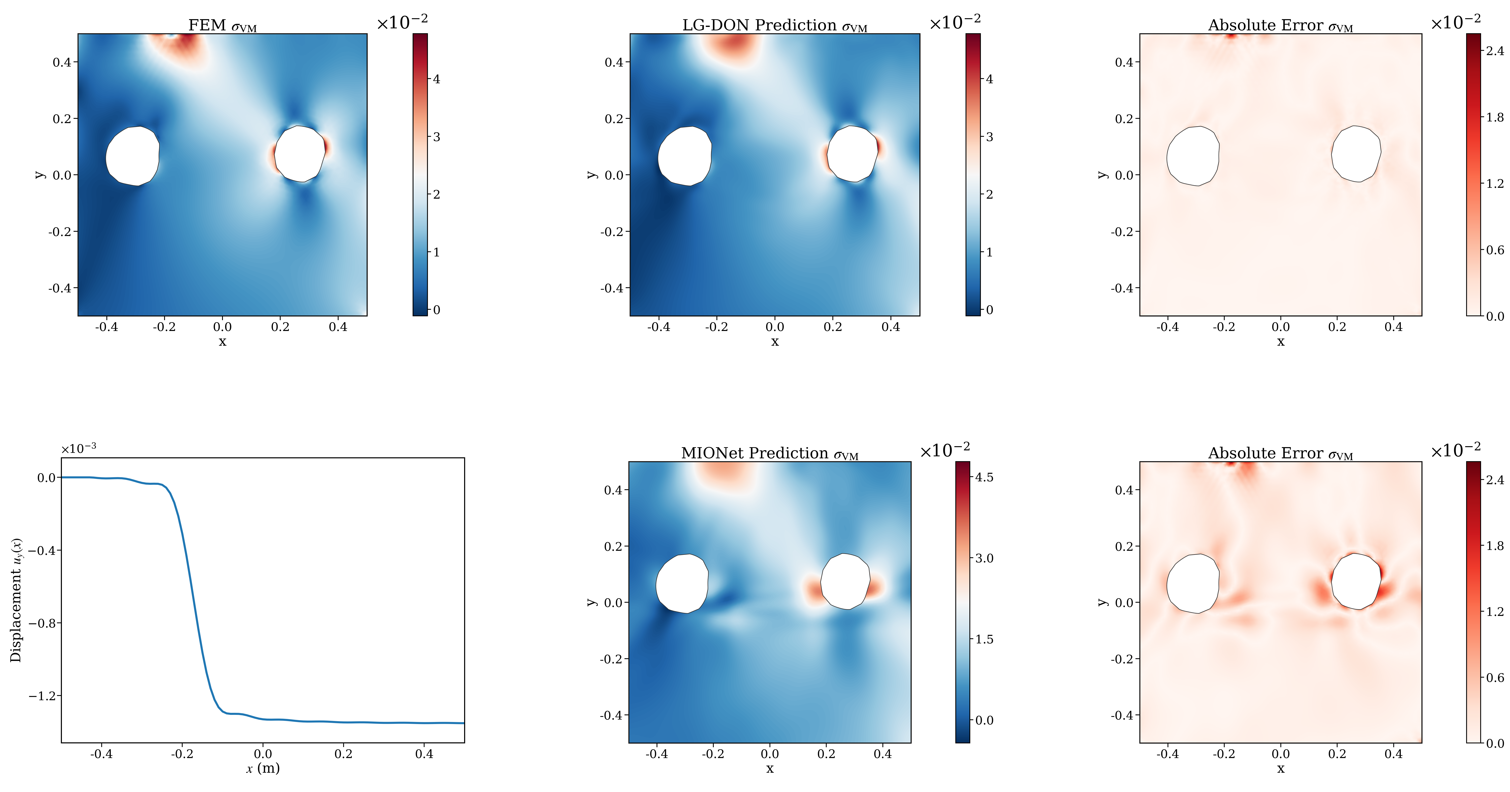}
\caption{Comparison of LG-DON and modified MIONet predictions under localized imposed-displacement excitation.}
\label{fig:localized_displacement}
\end{figure}

Fig.~\ref{fig:localized_displacement} compares the von Mises stress predictions of LG-DON and modified MIONet under localized prescribed displacement for the plate with two holes. Both models reproduced the overall stress distribution, but LG-DON predictions were closer to the FEM reference solution in regions with steep gradients around the irregular holes.

\begin{figure}[pos=!htbp]
\centering
\includegraphics[width=\linewidth]{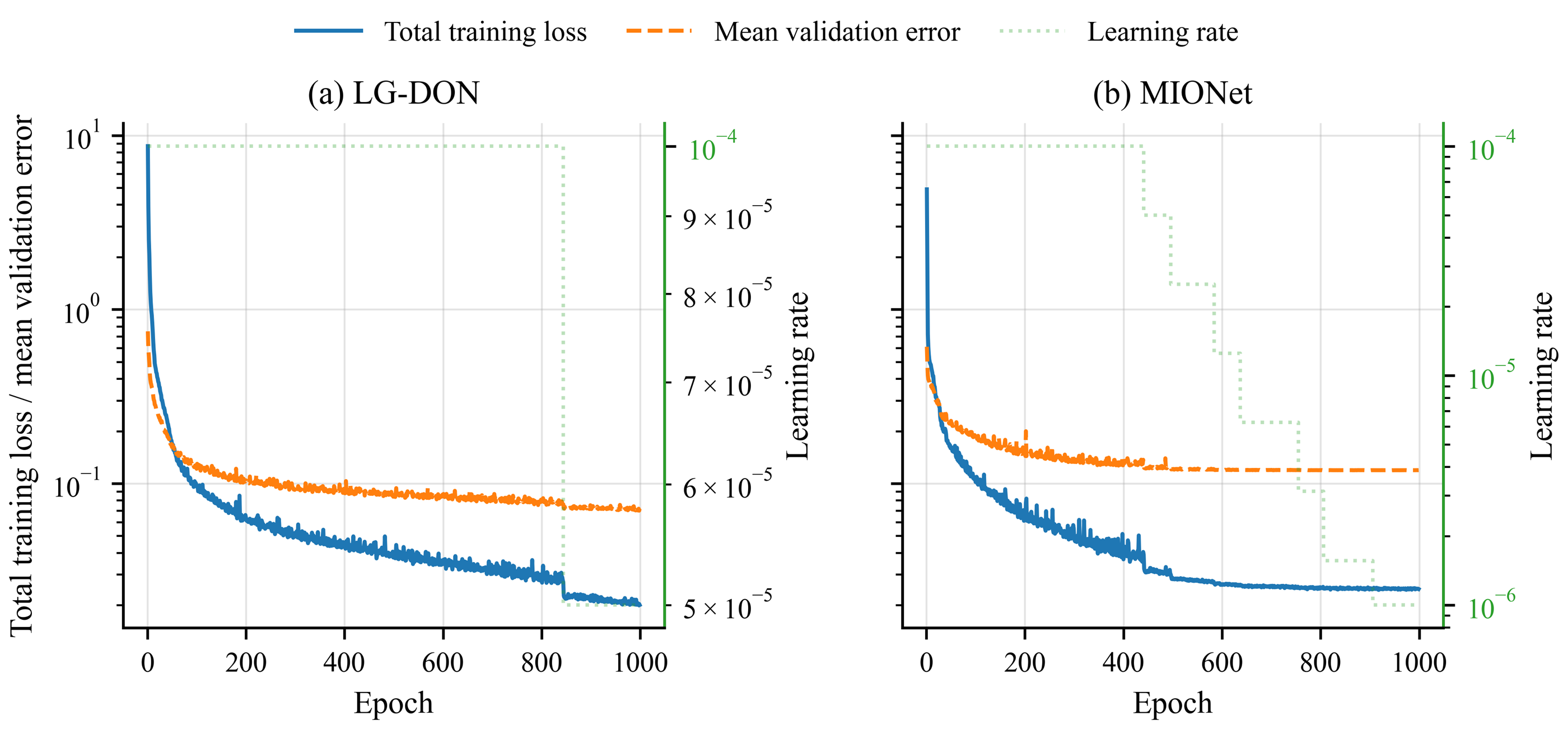}
\caption{Training convergence of LG-DON and modified MIONet: total training loss, mean validation error, and learning-rate schedule.}
\label{fig:training_convergence}
\end{figure}

Fig.~\ref{fig:training_convergence} compares the training behavior of the two models in terms of total training loss, mean validation relative error, and learning rate. For both models, the training loss decreased rapidly during the initial stage and then gradually stabilized. The mean validation error of LG-DON continued to decrease throughout training, whereas that of modified MIONet remained higher and approached a plateau at later epochs. Accordingly, the learning rate of LG-DON was reduced only once, late in training. By comparison, the learning rate of modified MIONet was reduced several times after the validation metric ceased to improve. These trends suggest that modified MIONet reached a performance plateau earlier within the 1000-epoch training period. Under the same training data and optimization strategy, the curves are consistent with the lower overall validation error achieved by LG-DON when boundary inputs and geometry vary jointly.

The quantitative results on the complete test set are summarized in table.~\ref{tab:plate_errors}. The relative $L_2$ errors of LG-DON for $u_x$, $u_y$, and $\sigma_{vm}$ were 6.26\%, 2.71\%, and 13.26\%, respectively, compared with 4.28\%, 2.92\%, and 29.30\% for modified MIONet. Although modified MIONet predicted $u_x$ more accurately, LG-DON achieved a slightly lower error for $u_y$ and a substantially lower von Mises stress error, resulting in a lower overall mean relative $L_2$ error.

\begin{table}[pos=!htbp]
\centering
\caption{Relative $L_2$ errors of LG-DON and modified MIONet over the complete test set across four loading scenarios.}
\label{tab:plate_errors}
\renewcommand{\arraystretch}{1.15}
\setlength{\tabcolsep}{5pt}

\begin{tabularx}{\linewidth}{@{}>{\raggedright\arraybackslash}X*{3}{>{\raggedleft\arraybackslash}p{0.105\linewidth}}>{\raggedleft\arraybackslash}p{0.18\linewidth}@{}}
\toprule
Method & $u_x$ & $u_y$ & $\sigma_{vm}$ & mean relative $L_2$ \\
\midrule
LG-DON & 6.26\% & 2.71\% & 13.26\% & 7.41\% \\
Modified MIONet & 4.28\% & 2.92\% & 29.30\% & 12.17\% \\
\bottomrule
\end{tabularx}
\end{table}

\begin{figure}[pos=!htbp]
\centering
\includegraphics[width=0.95\linewidth]{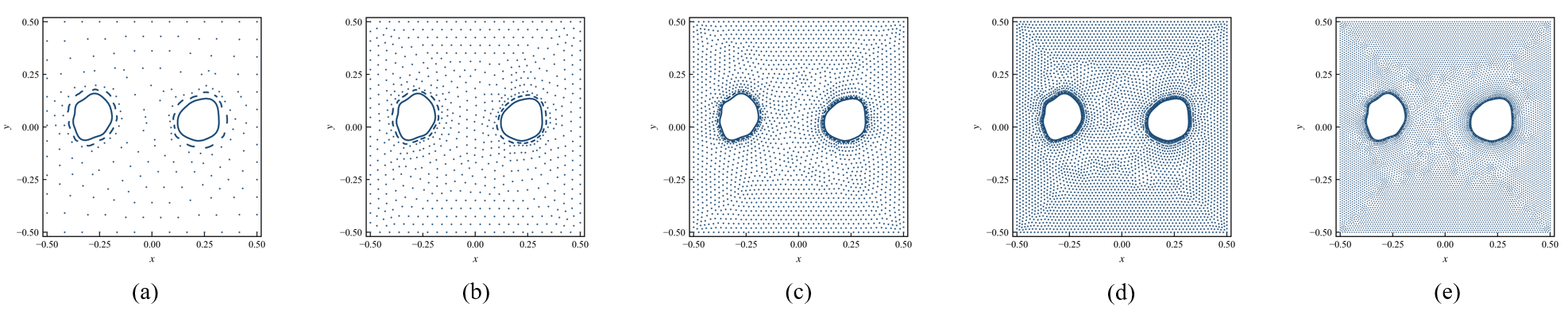}
\caption{Finite element discretizations with different numbers of output nodes: $N$ = 1000, 1500, 3000, 5000, and 10000.}
\label{fig:discretizations}
\end{figure}

To examine the effects of spatial discretization on the prediction accuracy of both models and on the individual response components learned by LG-DON, we constructed datasets with different numbers of spatial discretization points. As shown in Fig.~\ref{fig:discretizations}, five levels were considered: $N$ = 1000, 1500, 3000, 5000, and 10000. Here, $N$ denotes the number of discrete nodes used to represent the computational domain or response fields of each geometry sample. The original discretization used $N$ = 8000 nodes.

\begin{figure}[pos=!htbp]
\centering
\includegraphics[
  width=\linewidth,
  height=0.75\textheight,
  keepaspectratio
]{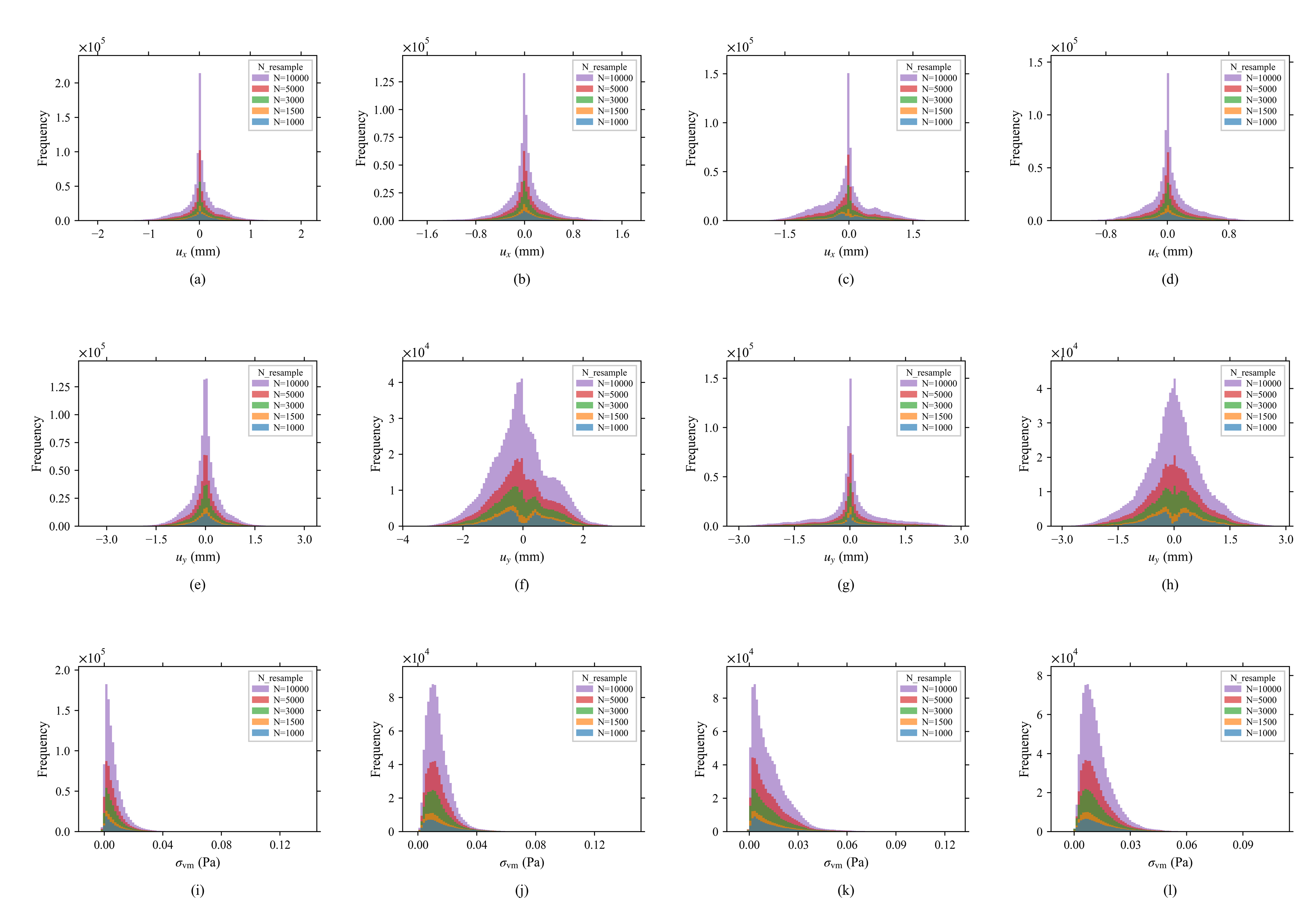}
\caption{Distributions of displacement components $u_x$, $u_y$ and von Mises stress $\sigma_{vm}$ predicted by LG-DON at different output discretization levels under four boundary input conditions. (a, e, i) Concentrated forces; (b, f, j) distributed traction; (c, g, k) localized prescribed displacement; (d, h, l) full-boundary prescribed displacement. The first, second, and third rows show $u_x$, $u_y$, and $\sigma_{vm}$, respectively.}
\label{fig:discretization_distributions}
\end{figure}

Fig.~\ref{fig:discretization_distributions} compares the predicted-value distributions of $u_x$, $u_y$, and $\sigma_{vm}$ across spatial discretization levels and boundary input conditions. Changes in the distribution shapes are most apparent for $\sigma_{vm}$ under imposed-displacement loading.

\begin{figure}[pos=!htbp]
\centering
\includegraphics[width=\linewidth]{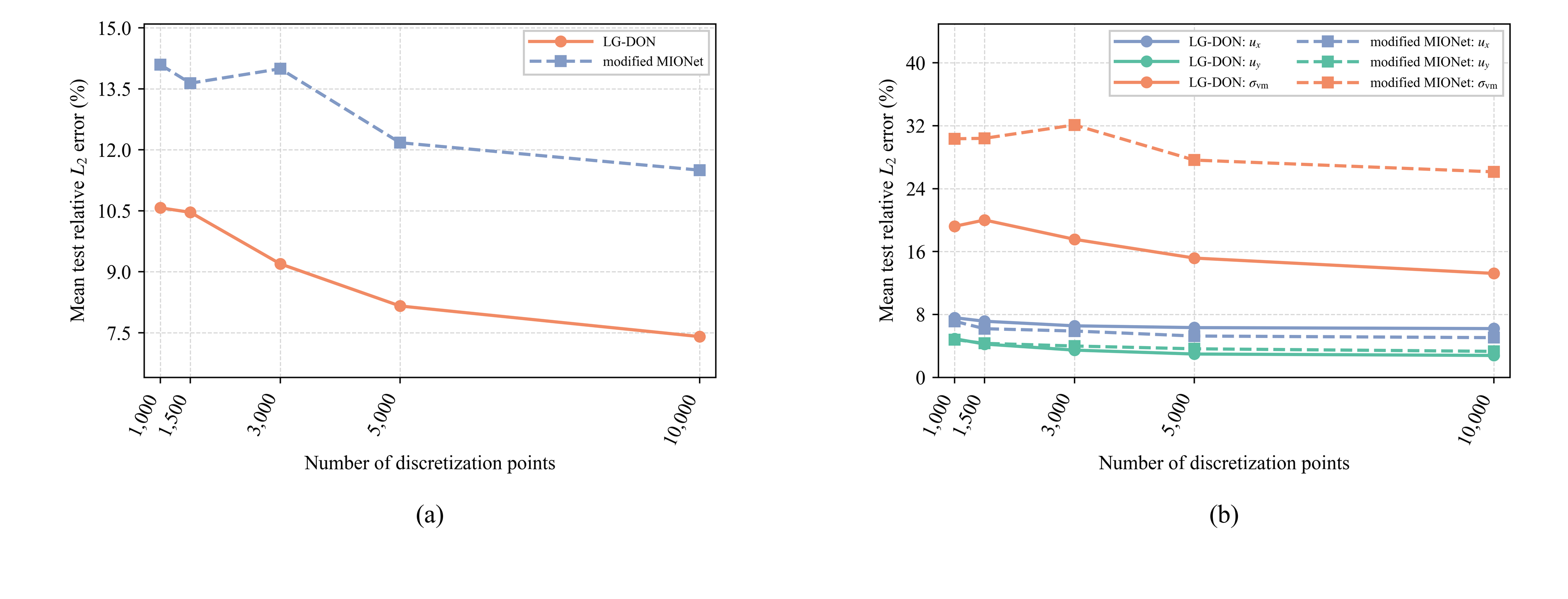}
\caption{Effect of output discretization on the prediction accuracy of LG-DON and modified MIONet: (a) overall mean test relative $L_2$ errors; (b) mean test relative $L_2$ errors of $u_x$, $u_y$, and $\sigma_{vm}$.}
\label{fig:discretization_accuracy}
\end{figure}

To quantify prediction accuracy at each discretization level and compare LG-DON with modified MIONet, Fig.~\ref{fig:discretization_accuracy} reports their mean relative $L_2$ errors. Fig.~\ref{fig:discretization_accuracy}(a) compares the overall errors, while Fig.~\ref{fig:discretization_accuracy}(b) presents the errors of the individual response components.

As shown in Fig.~\ref{fig:discretization_accuracy}(a), the overall error of LG-DON decreases monotonically with increasing N. Modified MIONet also shows an overall decreasing trend, but with slight nonmonotonic variations, reaching approximately 11.5\% at N=10000. LG-DON achieves a lower mean relative L2 error at all discretization levels and exhibits a more consistent improvement with increasing N. These results indicate that, when trained and tested independently at each discretization level, both models generally benefit from finer spatial discretization.

Fig.~\ref{fig:discretization_accuracy}(b) further compares the errors in $u_x$, $u_y$, and von Mises stress. For LG-DON, the $u_x$ error decreases from approximately 7.8\% to 6.3\%, while the $u_y$ error decreases from 5.0\% to 2.9\%, with only limited improvement beyond N=5000. Modified MIONet yields lower $u_x$ errors, whereas LG-DON achieves lower $u_y$ errors at higher discretization levels. The von Mises stress is more sensitive to discretization density. Its error for LG-DON decreases overall from approximately 19.3\% to 13.2\%, while that of modified MIONet reaches about 32\% near N=3000 and decreases to approximately 26\% at N=10000, remaining higher than LG-DON throughout.

Overall, increasing the number of discretization nodes improves the prediction accuracy of both models, although the displacement errors tend to level off at finer discretizations. The von Mises stress shows a stronger dependence on discretization density and represents the main performance difference between the two models. This may be attributed to the stronger local variations of the stress field near hole boundaries and loading regions, where finer discretization provides richer spatial information. Although modified MIONet performs better for $u_x$, the lower stress errors of LG-DON result in a lower overall prediction error.


\subsection{Cantilever Beam with an Irregular Hole}\label{subsec:cantilever}

\begin{figure}[pos=!htbp]
\centering
\includegraphics[width=\linewidth]{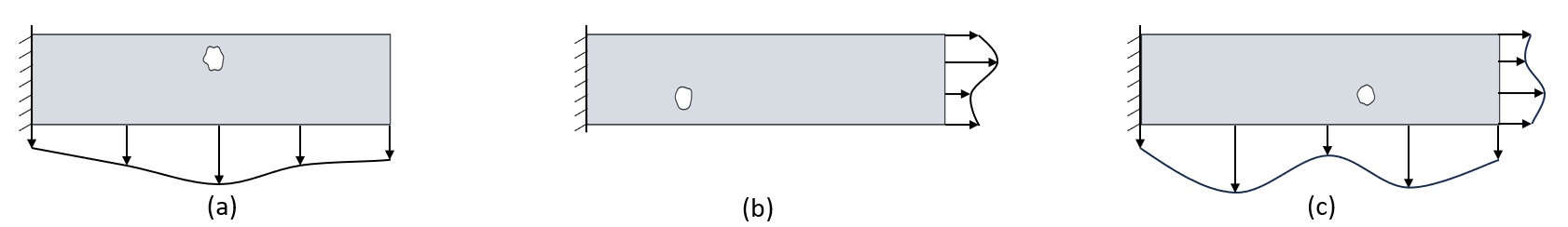}
\caption{Schematic illustration of the three loading scenarios for the irregular-hole cantilever beam: (a) distributed-traction loading, (b) imposed-displacement loading, and (c) combined distributed-traction and imposed-displacement loading.}
\label{fig:cantilever_scenarios}
\end{figure}

To further evaluate the structural-response prediction capability of LG-DON under varying internal geometries and multiple boundary excitations, this case considers the two-dimensional cantilever beam with a single irregular hole shown in Fig.~\ref{fig:cantilever_scenarios}. The computational domain is 2 m $\times$ 0.4 m, with the left boundary fully fixed. The material is assumed to be homogeneous, isotropic, and linearly elastic, with Young's modulus $E = 2 \times 10^{4}$ Pa and Poisson's ratio $v = 0.25$. The irregular hole is generated using Eq.~\eqref{eq:hole_boundary} with 32 boundary points sampled at equally spaced polar angles. In the training data, all radial values defining the hole boundary are constrained to $[0.03, 0.08]$ m, and the $x$-coordinate of the hole center, $c_x$, is randomly sampled from $[-0.5, 0.5]$ m. The $y$-coordinate, $c_y$, is determined dynamically according to the realized radial extent of the hole and the prescribed clearance from the outer boundary. Consequently, both the hole position and boundary shape vary across samples. For the boundary excitations, a spatially varying distributed traction is applied in the y direction along the bottom boundary, with a training amplitude range of 0.1--3.0 Pa, while a spatially varying imposed displacement is applied in the x direction along the right boundary, with a training amplitude range of $4.0 \times 10^{-5}$--$5.0 \times 10^{-3}$ m.

Based on these two types of boundary excitation, three loading scenarios are considered, as shown in Fig.~\ref{fig:cantilever_scenarios}: Scenario 1 includes only the bottom distributed traction; Scenario 2 includes only the imposed displacement on the right boundary; and Scenario 3 combines both boundary excitations.

\begin{figure}[pos=!htbp]
\centering
\includegraphics[width=\linewidth]{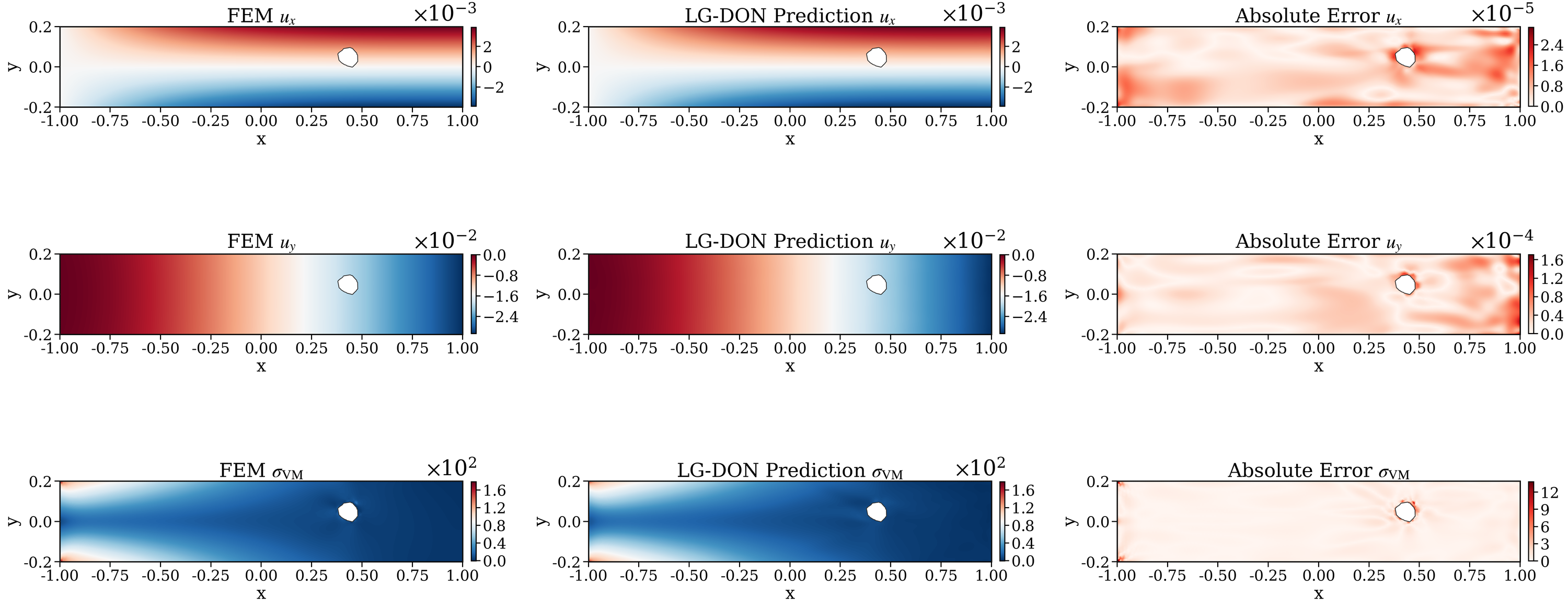}
\caption{In-distribution predictions under distributed-traction loading (Scenario 1).}
\label{fig:cantilever_id}
\end{figure}

After training, all network parameters were fixed, and the in-distribution (ID) prediction performance was first evaluated on a test set sampled from the same parameter distributions as the training data. Fig.~\ref{fig:cantilever_id} shows representative prediction results for Scenario 1. LG-DON closely reproduces the overall displacement-field distributions obtained by FEM and captures the local stress concentrations near the fixed end and the irregular hole. The prediction errors are concentrated mainly along the hole boundary and in other local regions with large response gradients, whereas the predictions over the main body of the structure remain in good agreement with the FEM reference solution. Together with the test-set statistics for all three loading scenarios in table.~\ref{tab:cantilever_accuracy}, these results show that LG-DON maintains relatively low mean relative $L_2$ errors within the original load and geometry parameter ranges, indicating that the model can learn structural response mappings associated with different boundary excitations and varying geometries within a unified framework.

\begin{figure}[pos=!htbp]
\centering
\includegraphics[width=\linewidth]{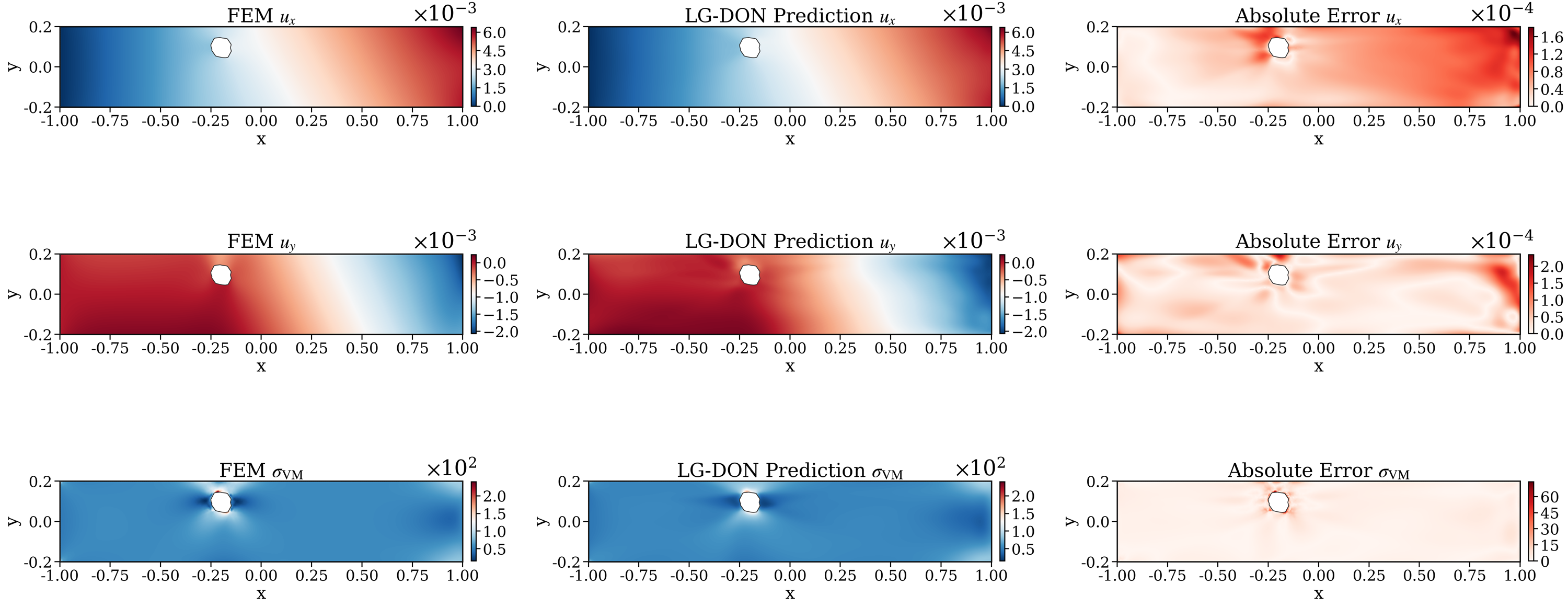}
\caption{Load extrapolation under imposed-displacement loading (Scenario 2).}
\label{fig:cantilever_load_ood}
\end{figure}

\begin{figure}[pos=!htbp]
\centering
\includegraphics[width=\linewidth]{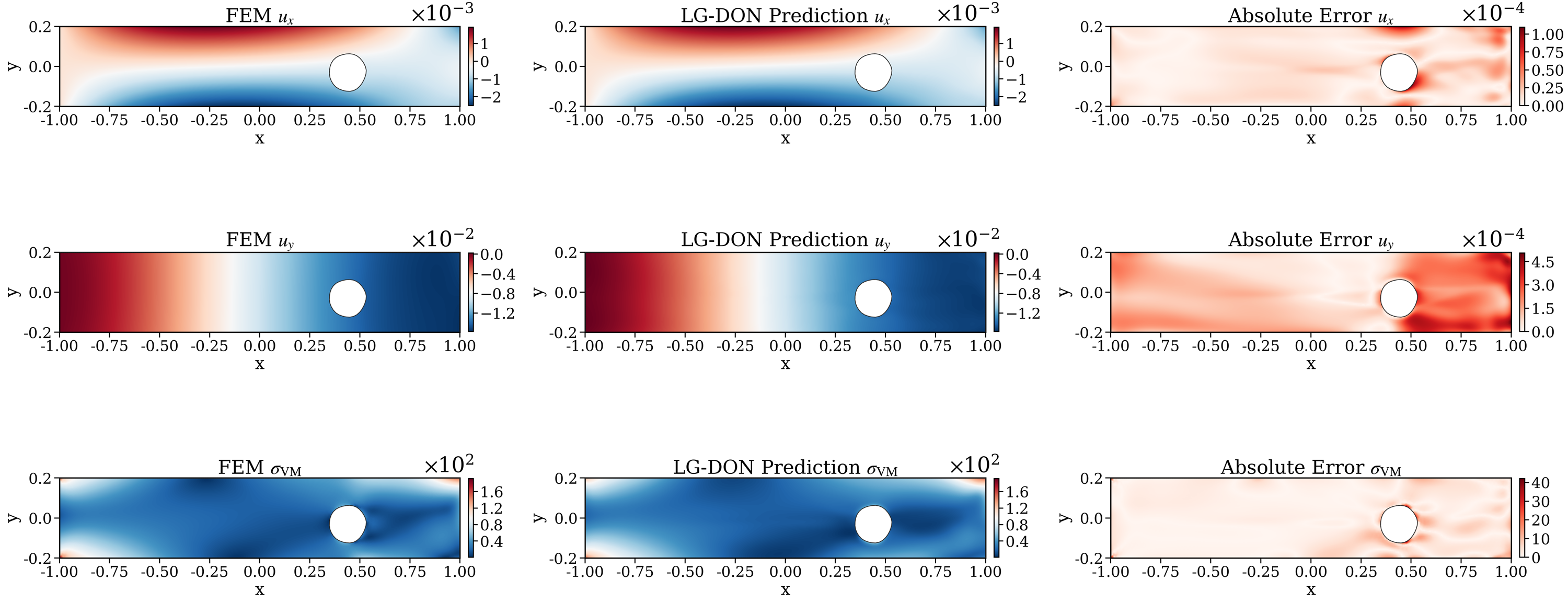}
\caption{Geometry-parameter extrapolation under combined loading (Scenario 3).}
\label{fig:cantilever_geom_ood}
\end{figure}

To further assess out-of-distribution (OOD) generalization beyond the training parameter ranges, separate load-amplitude and geometry extrapolation tests were conducted. Fig.~\ref{fig:cantilever_load_ood} presents representative load-amplitude extrapolation results for Scenario 2. In this test, the geometry distribution is kept identical to that used for training, while the amplitude of the imposed displacement on the right boundary is shifted from the training range of $4.0 \times 10^{-5}$--$5.0 \times 10^{-3}$ m to an extrapolation range of $5.2 \times 10^{-3}$--$7.2 \times 10^{-3}$ m. For the load-extrapolation test sets of the scenarios involving bottom traction, the traction amplitude is likewise shifted from 0.1--3.0 Pa to 3.2--4.2 Pa. For all extrapolation tests, the normalization parameters determined from the training data were kept fixed and were not recomputed using the out-of-distribution test data. Fig.~\ref{fig:cantilever_geom_ood} presents representative geometry-extrapolation results for Scenario 3. In this test, the boundary excitations remain within their original training distributions, while the radial range of the irregular hole is shifted from $[0.03, 0.08]$ m during training to $[0.085, 0.10]$ m. Under both extrapolation settings, LG-DON preserves the principal displacement-response patterns observed in the FEM reference solutions. The errors remain localized mainly near the irregular hole and the loaded boundaries, where the response gradients are relatively large. The field-wise errors exhibit a clear scenario dependence: $\sigma_{vm}$ shows the largest OOD error in Scenarios 1 and 3, whereas $u_y$ is the most sensitive response component in Scenario 2. The quantitative results in table.~\ref{tab:cantilever_accuracy} further show that both load-amplitude extrapolation and geometry-parameter extrapolation increase the prediction error and sample-to-sample variability relative to the ID test set. For Scenarios 1--3, the mean relative $L_2$ errors increase from 2.17\%, 5.85\%, and 4.18\% in the ID setting to 5.02\%, 12.88\%, and 7.55\% under load-amplitude extrapolation, and to 5.85\%, 16.20\%, and 9.14\% under geometry-parameter extrapolation, respectively. At the field level, the largest extrapolation errors occur in $\sigma_{vm}$ for Scenarios 1 and 3, whereas Scenario 2 exhibits a substantially larger error in $u_y$, reaching 27.18\% under load-amplitude extrapolation and 25.17\% under geometry-parameter extrapolation.

\begin{table}[pos=!htbp]
\centering
\caption{Accuracy comparison of LG-DON under in-distribution, load-amplitude-extrapolation, and geometry-parameter extrapolation settings.}
\label{tab:cantilever_accuracy}

\footnotesize
\setlength{\tabcolsep}{3.5pt}
\renewcommand{\arraystretch}{1.20}

\begin{tabularx}{\linewidth}{@{}
  >{\raggedright\arraybackslash}p{0.18\linewidth}
  >{\raggedright\arraybackslash}p{0.14\linewidth}
  *{3}{>{\raggedleft\arraybackslash}p{0.08\linewidth}}
  >{\raggedleft\arraybackslash}X
  >{\raggedleft\arraybackslash}p{0.145\linewidth}
@{}}
\toprule
Evaluation setting
& Loading scenario
& $u_x$
& $u_y$
& $\sigma_{vm}$
& \shortstack[r]{Mean relative\\$L_2$ error}
& \shortstack[r]{Standard\\deviation}
\\
\midrule

\multirow{3}{=}{In-distribution}
& Scenario 1 & 0.92\% & 0.60\% & 4.99\% & 2.17\% & 1.52\% \\
& Scenario 2 & 0.88\% & 6.46\% & 10.21\% & 5.85\% & 4.56\% \\
& Scenario 3 & 1.09\% & 1.59\% & 9.87\% & 4.18\% & 1.76\% \\
\midrule

\multirow{3}{=}{Load-amplitude extrapolation}
& Scenario 1 & 4.66\% & 4.18\% & 6.21\% & 5.02\% & 1.65\% \\
& Scenario 2 & 3.06\% & 27.18\% & 8.39\% & 12.88\% & 7.73\% \\
& Scenario 3 & 4.65\% & 4.26\% & 13.75\% & 7.55\% & 1.96\% \\
\midrule

\multirow{3}{=}{Geometry extrapolation}
& Scenario 1 & 4.70\% & 3.46\% & 9.40\% & 5.85\% & 4.29\% \\
& Scenario 2 & 4.37\% & 25.17\% & 19.07\% & 16.20\% & 12.47\% \\
& Scenario 3 & 3.92\% & 6.25\% & 17.24\% & 9.14\% & 4.10\% \\
\bottomrule
\end{tabularx}
\end{table}

Taken together, Figs.~\ref{fig:cantilever_id}, \ref{fig:cantilever_load_ood}, and~\ref{fig:cantilever_geom_ood} and table.~\ref{tab:cantilever_accuracy} show that a single LG-DON model can handle variations in irregular-hole position and shape under distributed traction, imposed displacement, and combined loading. Although the errors increase under both load-amplitude and geometry extrapolation, the principal response patterns remain consistent with the FEM reference solutions. Within the irregular-hole geometry family considered here, LG-DON therefore exhibits a degree of both load-amplitude and geometry-parameter extrapolation, although the sensitivity of individual response fields to OOD inputs is scenario dependent.


\subsection{L-Shaped Bracket with Varying External Geometry}\label{subsec:bracket}

\begin{figure}[pos=!htbp]
\centering
\includegraphics[width=0.95\linewidth]{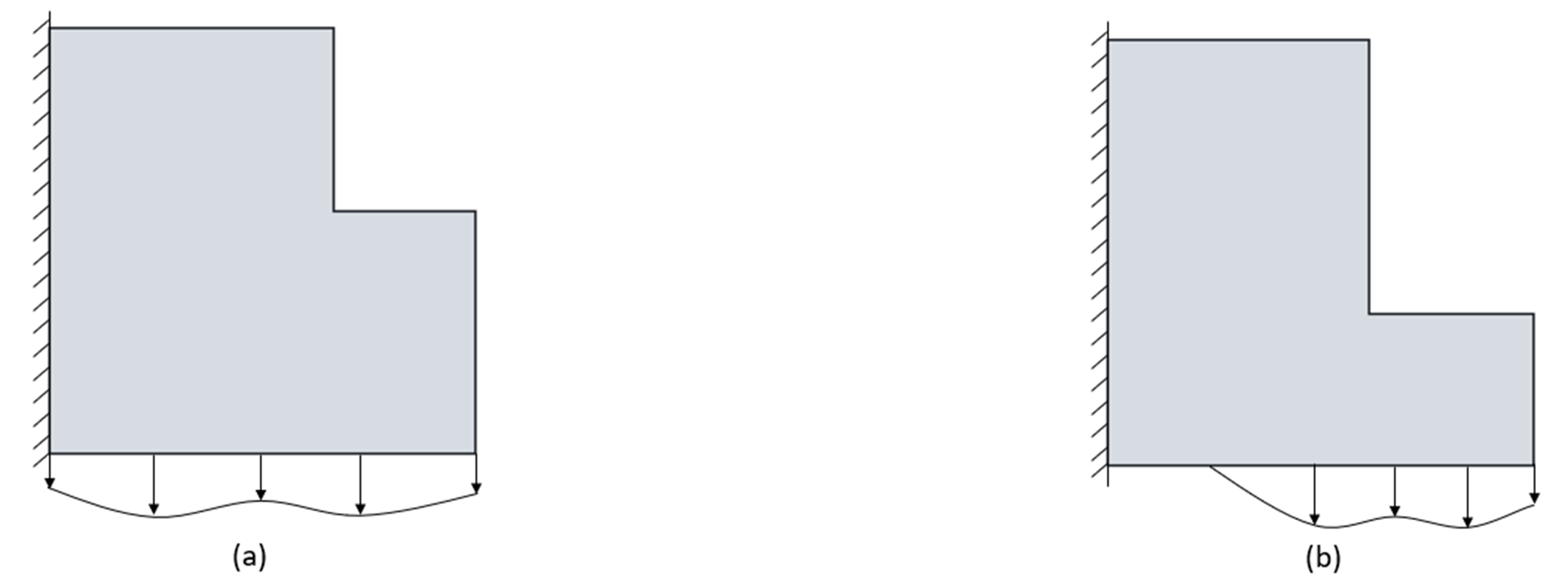}
\caption{Schematic illustration of the L-shaped bracket with varying external geometry under two loading scenarios: (a) distributed-traction loading and (b) localized imposed-displacement loading.}
\label{fig:bracket_scenarios}
\end{figure}

The first two cases focus on varying internal boundaries within a fixed external computational domain. To further investigate adaptation to external-geometry variation, this case considers an L-shaped bracket in which changes to the external boundary simultaneously alter the material domain, boundary locations, and the locations of local high-gradient response regions. As shown in Fig.~\ref{fig:bracket_scenarios}, the initial computational domain is a square with side length 1 m. An L-shaped bracket is formed by removing a rectangular region of size $L_x \times L_y$ from the upper-right corner, where $L_x$ and $L_y$ are independently sampled from $[0.25,0.65]$ m. Thus, both the re-entrant corner location and the material domain vary across samples. The material is homogeneous, isotropic, and linearly elastic, with $E = 2 \times 10^4$ Pa and $v = 0.25$. The left boundary is fixed, and two types of excitation are applied along the bottom boundary: in Fig.~\ref{fig:bracket_scenarios}(a), a spatially varying y-direction traction is applied over the entire bottom boundary; in Fig.~\ref{fig:bracket_scenarios}(b), a localized y-direction imposed displacement is applied over the right portion of the bottom boundary. For the localized displacement, a cubic smoothstep transition is used over $x \in [-0.25,0]$ m, and the full imposed-displacement amplitude is reached to the right of this interval.

\begin{figure}[pos=!htbp]
\centering
\includegraphics[width=0.95\linewidth]{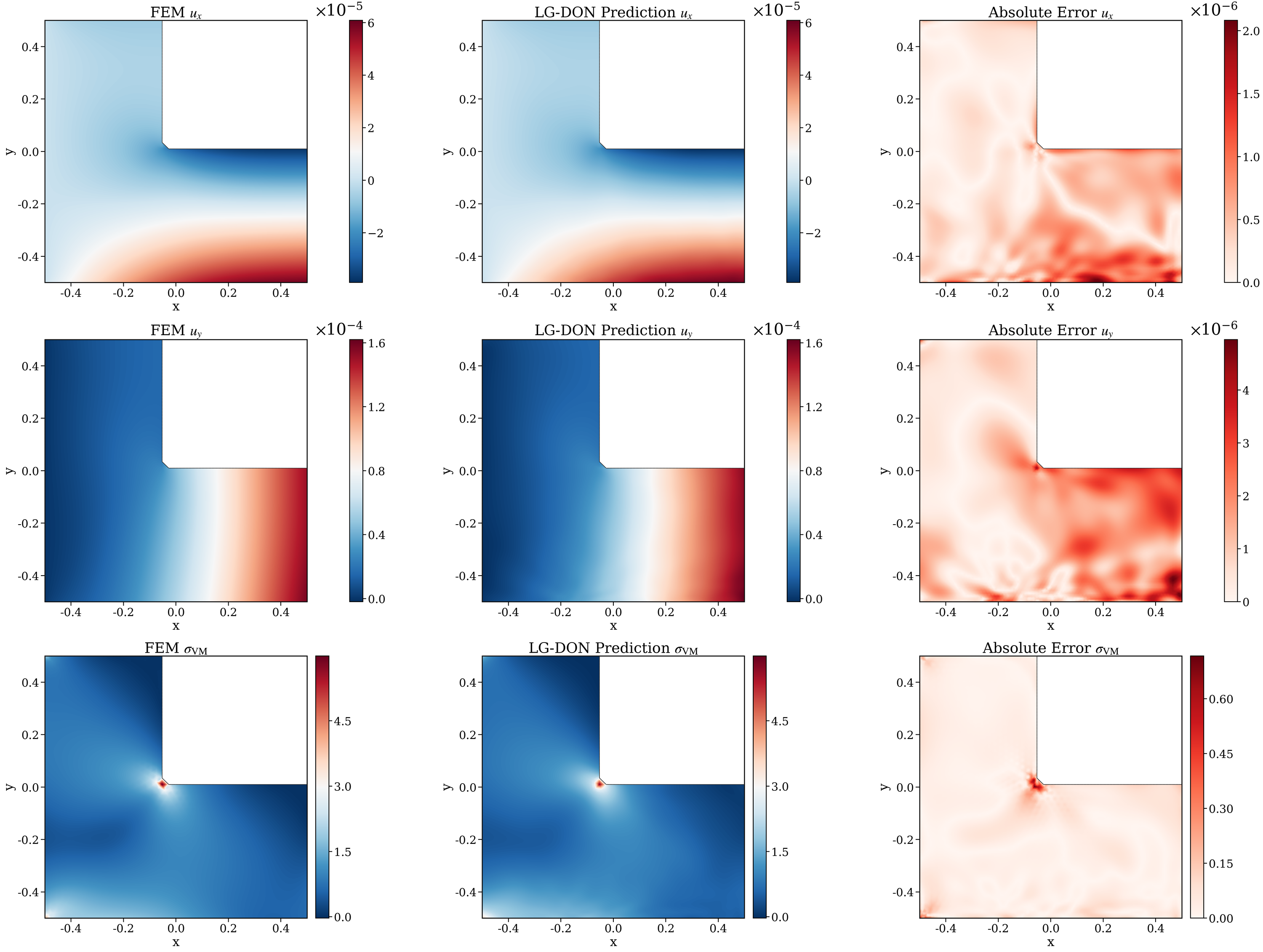}
\caption{Prediction results of LG-DON for the L-shaped bracket under distributed-traction loading.}
\label{fig:bracket_traction}
\end{figure}

\begin{figure}[pos=!htbp]
\centering
\includegraphics[width=0.95\linewidth]{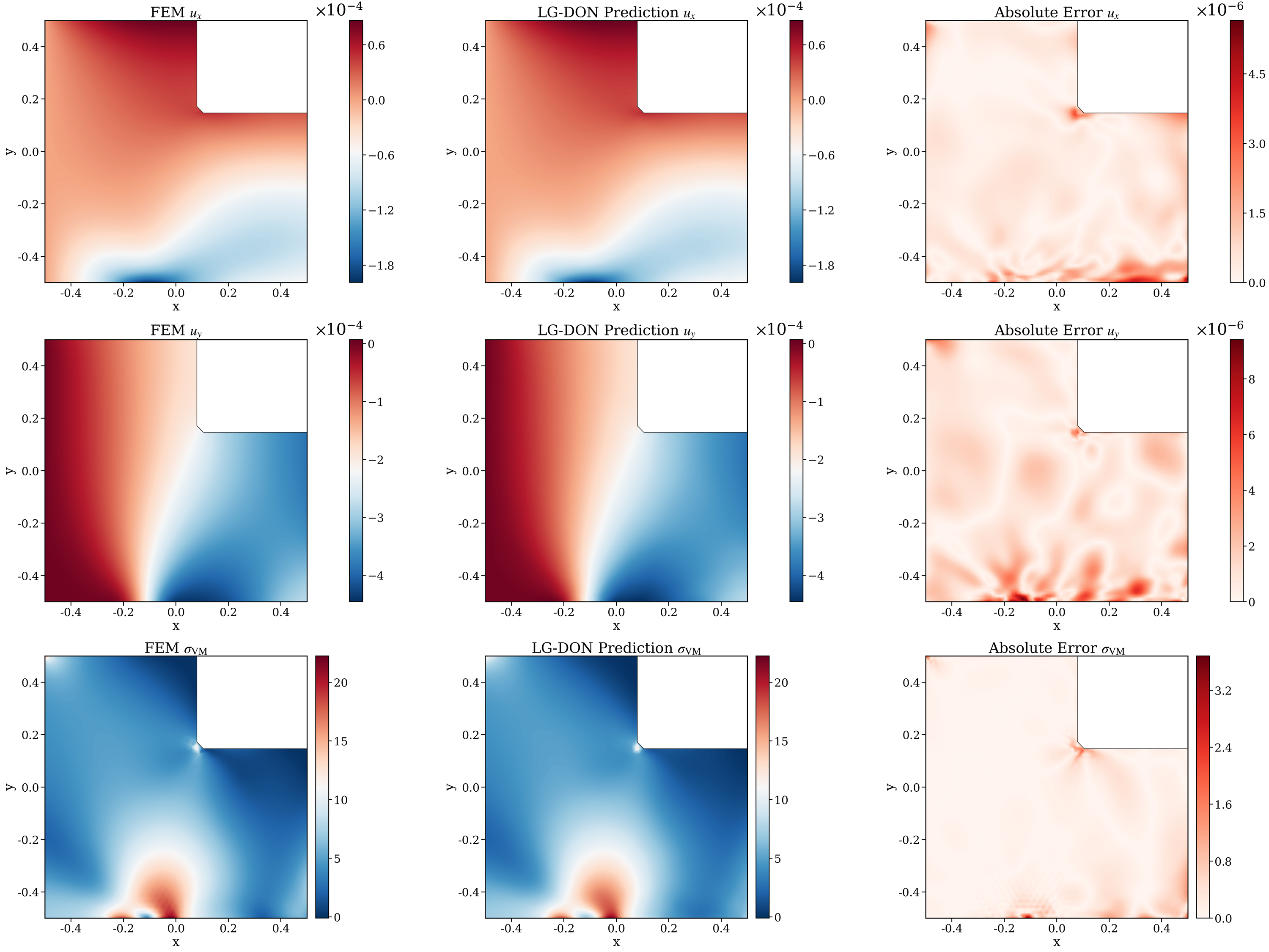}
\caption{Prediction results of LG-DON for the L-shaped bracket under localized imposed-displacement loading.}
\label{fig:bracket_displacement}
\end{figure}

Fig.~\ref{fig:bracket_traction} and Fig.~\ref{fig:bracket_displacement} show representative prediction results under distributed traction and localized imposed displacement, respectively. Because the two test samples have different $L_x$ and $L_y$ values, their re-entrant corner locations and effective material domains differ. Despite simultaneous variations in geometry and boundary excitation, LG-DON reproduces the overall distributions of the displacement and von Mises stress fields and identifies the local stress-concentration regions that shift with the re-entrant corner. This observation suggests that for the cases considered here, the predictions are not tied to fixed-location response patterns; rather, the load--geometry-response mapping adapts to the current geometry and loading condition. The absolute errors are concentrated mainly near the re-entrant corner, the bottom loading boundary, and the region of localized imposed displacement, where the response gradients are high, whereas the predictions over the main body of the structure remain consistent with the FEM reference. Taken together, the results for the two external geometries and loading types indicate that the geometry-adaptation capability of LG-DON is not limited to variations of internal holes within a fixed external boundary. The trained model can also directly predict structural responses when both the external geometry and loading condition vary, without retraining or transfer learning.

\section{Effectiveness of geometry representations and ablation analysis of joint encoding}\label{sec:ablation}

To evaluate the effectiveness of the geometry representations used in LG-DON, individually and in combination, we conducted geometry-encoding ablation studies on the three cases in Section 3: the plate with two holes, the cantilever beam with one hole, and the L-shaped bracket. We first examined the effects of the DG input point set and the SDF-based boundary-distance features separately. We then evaluated the incremental benefits, conditional complementarity, and cross-case consistency arising from their combination. Each configuration was trained independently three times. The overall performance metric for each run was the mean relative $L_2$ error on the test set, defined in Eq.~\eqref{eq:relative_l2_error} of Section 3. Each configuration was evaluated using the average of this overall test error over the three independent runs:

\begin{equation}
\label{eq:ablation_mean_error}
\overline{E}
=
\frac{1}{3}
\sum_{r=1}^{3} E_{mean,r}
\end{equation}

\subsection{Effectiveness of individual geometry representations}\label{subsec:single_encoding}

To examine the effect of the DG input point set, we first disabled the SDF-based boundary-distance features and retained only DG. Three point-set configurations were compared: all\_domain, all\_bound, and vary\_bound. These correspond to `Interior and boundary points', `Boundary points (fixed and variable)', and `Boundary points (variable only)' in table.~\ref{tab:dg_point_sets}, respectively. All three cases exhibited the same ranking: vary\_bound achieved the lowest average test error, followed by all\_bound and then all\_domain. This consistency suggests that, for the varying geometries considered here, the effectiveness of DG depends more on how well the input points distinguish geometric differences than on simply increasing the number of geometry points. One possible explanation is that many interior points vary little in position across geometry samples. Aggregating them together with boundary points may therefore dilute the boundary information that characterizes geometric variation, consistent with the discussion of redundant interior points in \cite{zhong2025lgdon19}.

\begin{table}[pos=!htbp]
\centering
\caption{Mean relative $L_2$ error for different DG point-set configurations.}
\label{tab:dg_point_sets}

\footnotesize
\setlength{\tabcolsep}{3pt}
\renewcommand{\arraystretch}{1.18}

\begin{tabularx}{\linewidth}{@{}>{\raggedright\arraybackslash}X*{3}{>{\raggedleft\arraybackslash}p{0.23\linewidth}}@{}}
\toprule
case
&
\shortstack{Interior and\\boundary points}
&
\shortstack{Boundary points\\(fixed and variable)}
&
\shortstack{Boundary points\\(variable only)}
\\
\midrule

Plate with two holes
& 9.50\% & 8.72\% & 8.63\% \\

Cantilever beam with one hole
& 8.01\% & 4.34\% & 4.30\% \\

L-shaped bracket
& 9.35\% & 8.99\% & 8.65\% \\

\bottomrule
\end{tabularx}
\end{table}

\begin{table}[pos=!htbp]
\centering
\caption{Mean relative $L_2$ error for different SDF configurations.}
\label{tab:sdf_configurations}

\renewcommand{\arraystretch}{1.15}
\setlength{\tabcolsep}{5pt}

\begin{tabularx}{\linewidth}{@{}>{\raggedright\arraybackslash}X*{2}{>{\raggedleft\arraybackslash}p{0.22\linewidth}}@{}}
\toprule
case & outer & boundary \\
\midrule

Plate with two holes
& 16.41\% & 14.95\% \\

Cantilever beam with one hole
& 7.92\% & 7.75\% \\

L-shaped bracket
& 29.91\% & 11.99\% \\

\bottomrule
\end{tabularx}

\par\vspace{3pt}
{\raggedright\footnotesize
Note: Both configurations use the complete domain boundary. Outer denotes the minimum distance across all boundary categories, whereas boundary denotes the vector of category-specific distances.
\par}
\end{table}

We next disabled DG and retained only the SDF-based boundary-distance features to compare the outer and boundary representations. As shown in table.~\ref{tab:sdf_configurations}, boundary achieved a lower average test error than outer in all three cases: the error decreased from 16.41\% to 14.95\% for the plate with two holes, from 7.92\% to 7.75\% for the cantilever beam with one hole, and from 29.91\% to 11.99\% for the L-shaped bracket. Since both representations cover the same complete boundary, these results suggest that retaining separate boundary distances is more effective for prediction than using only the nearest-boundary distance under the conditions considered. A possible explanation is that the separate channels identify the corresponding boundary categories and retain distances to other boundaries, providing spatial information that is lost when only the minimum distance is used.

The benefit of boundary over outer nevertheless differed considerably across the three cases, indicating that the performance of the SDF representation alone is already case-dependent. Overall, the experiments in Section 4.1 identify consistent configuration preferences when each geometry representation is used alone. They do not, however, directly establish the best joint configuration, because interactions between DG and SDF may depend on their specific combination.


\subsection{Conditional complementarity and selection of joint geometry encodings}\label{subsec:joint_encoding}

Building on the experiments with individual geometry representations, we examined the complementarity between DG and the SDF-based boundary-distance features and its dependence on the encoding configuration and geometry case. Each of the three DG point sets was combined with either the category-resolved boundary distances or their scalar minimum, denoted by boundary and outer, respectively, yielding six joint DG--SDF configurations. As in the preceding experiments, each configuration was trained independently three times, and the overall test-set mean relative $L_2$ error was averaged over the three runs. Each joint configuration was compared with its matched DG-only and SDF-only baselines. This comparison quantifies the incremental benefit of adding one type of geometry information while keeping the other representation unchanged.

The percentage improvements of a joint configuration over its matched DG-only and SDF-only configurations are defined as

\begin{equation}
\label{eq:improvement_dg}
I_{DG}
=
\frac{
\overline{E}_{DG}
-
\overline{E}_{joint}
}{
\overline{E}_{DG}
}
\times 100\%
\end{equation}

\begin{equation}
\label{eq:improvement_sdf}
I_{SDF}
=
\frac{
\overline{E}_{SDF}
-
\overline{E}_{joint}
}{
\overline{E}_{SDF}
}
\times 100\%
\end{equation}

A positive improvement indicates that joint encoding reduces the prediction error; a negative value indicates that the joint configuration does not improve upon the corresponding single-encoding baseline. Table~\ref{tab:joint_configurations} reports the average test errors of the six joint DG--SDF configurations for the three cases and the mean improvements over the matched DG-only and SDF-only baselines in its final two columns. The two cases dominated by variations in internal hole boundaries exhibit similar configuration preferences. For both the plate with two holes and the cantilever beam with one hole, vary\_bound+boundary achieves the lowest average errors, at 7.28\% and 4.18\%, respectively. The all\_bound+boundary configuration ranks second in both cases, with errors of 7.35\% and 4.22\%, whereas the two configurations using all\_domain generally yield higher errors. In contrast, the L-shaped bracket, which involves variations in the outer contour, shows a different ranking: all\_bound+boundary achieves the lowest average error of 8.47\%, while vary\_bound+boundary yields 8.70\%.

\begin{figure}[pos=!htbp]
\centering
\includegraphics[width=\linewidth]{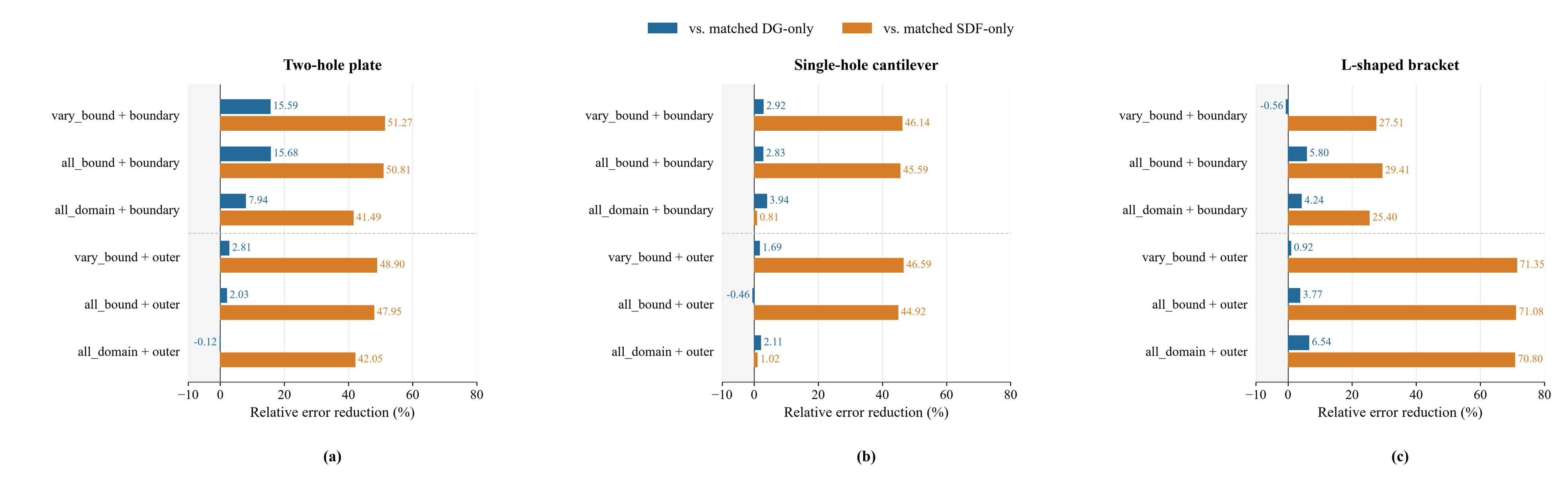}
\caption{Relative error reductions of joint DG--SDF configurations over matched single-encoding baselines: (a) two-hole plate; (b) single-hole cantilever; (c) L-shaped bracket.}
\label{fig:joint_encoding_improvements}
\end{figure}

For a fixed DG point set, boundary consistently produces lower average errors than outer for the plate and cantilever cases. For the L-shaped bracket, however, outer slightly outperforms boundary when combined with vary\_bound or all\_domain. These results indicate that the relative ranking of the joint DG--SDF configurations is not fully consistent across the three cases and that the advantage of a geometry representation used alone cannot be directly extrapolated to joint encoding. Fig.~\ref{fig:joint_encoding_improvements} further compares each joint configuration with its matched DG-only and SDF-only baselines to quantify the incremental benefits for each case. The table and figure together characterize both the final predictive performance of the joint configurations and the complementarity between the two types of geometry information.

\begin{table}[pos=!htbp]
\centering
\caption{Comparison of joint DG--SDF configurations across the three cases}
\label{tab:joint_configurations}

\scriptsize
\setlength{\tabcolsep}{2pt}
\renewcommand{\arraystretch}{1.18}

\begin{tabularx}{\linewidth}{@{}>{\raggedright\arraybackslash}p{0.205\linewidth}*{3}{>{\centering\arraybackslash}p{0.11\linewidth}}>{\centering\arraybackslash}X>{\centering\arraybackslash}X@{}}
\toprule
Joint configuration
&
\shortstack{Two-hole\\plate}
&
\shortstack{Single-hole\\cantilever}
&
\shortstack{L-shaped\\bracket}
&
\shortstack{Mean improvement\\across cases\\(vs. DG-only)}
&
\shortstack{Mean improvement\\across cases\\(vs. SDF-only)}
\\
\midrule

vary\_bound + boundary
& 7.28\% & 4.18\% & 8.70\% & 5.98\% & 41.64\% \\

vary\_bound + outer
& 8.39\% & 4.23\% & 8.57\% & 1.81\% & 55.62\% \\
\addlinespace[2pt]

all\_bound + boundary
& 7.35\% & 4.22\% & 8.47\% & 8.10\% & 41.94\% \\

all\_bound + outer
& 8.54\% & 4.36\% & 8.65\% & 1.78\% & 54.65\% \\
\addlinespace[2pt]

all\_domain + boundary
& 8.75\% & 7.69\% & 8.95\% & 5.37\% & 22.57\% \\

all\_domain + outer
& 9.51\% & 7.84\% & 8.73\% & 2.84\% & 37.96\% \\

\bottomrule
\end{tabularx}
\end{table}

Panels (a)--(c) of Fig.~\ref{fig:joint_encoding_improvements} correspond to the plate with two holes, the cantilever beam with one hole, and the L-shaped bracket, respectively. In each panel, the blue bars use the matched DG-only model as the baseline and show the improvement obtained by adding SDF features while keeping the DG point set unchanged. The orange bars use the matched SDF-only model as the baseline and show the improvement obtained by adding DG while keeping the SDF configuration unchanged. The first three configurations from top to bottom use boundary, whereas the last three use outer.

Comparison of the blue bars in Fig.~\ref{fig:joint_encoding_improvements}(a)--(c) shows that the incremental benefit of adding SDF to an existing DG representation depends on both the case and the configuration. For the two-hole plate, vary\_bound+boundary and all\_bound+boundary improve upon their matched DG-only baselines by 15.59\% and 15.68\%, respectively; both improvements exceed those obtained with the corresponding outer configurations. Together with their lower final errors in Table~\ref{tab:joint_configurations}, this result suggests that category-resolved boundary distances effectively complement boundary-focused DG representations in this case. For the single-hole cantilever, most improvements are modest, at approximately 1\%--4\%, with all\_bound+outer yielding $-0.46\%$. For the L-shaped bracket, vary\_bound+boundary yields $-0.56\%$, whereas all\_bound+boundary and all\_domain+outer achieve improvements of 5.80\% and 6.54\%, respectively; however, the former still produces the lower final average error.

The orange bars in Fig.~\ref{fig:joint_encoding_improvements} show that all joint configurations improve upon their matched SDF-only baselines. The improvements are approximately 41\%--51\% for the two-hole plate. For the single-hole cantilever, the improvements are approximately 45\%--47\% when SDF is combined with boundary-based DG point sets, but only about 1\% when combined with all\_domain. For the L-shaped bracket, the improvements are approximately 25\%--29\% for the boundary configurations and about 71\% for the outer configurations. In particular, the error of the outer-only configuration for the L-shaped bracket is 29.91\%, which decreases to 8.57\%--8.73\% after introducing DG. This large relative improvement is also associated with the relatively high SDF-only baseline error, while all\_bound+boundary still achieves the lowest final average error.

As summarized in Table~\ref{tab:joint_configurations}, the six joint configurations achieve mean improvements across the three cases of 1.78\%--8.10\% over DG-only and 22.57\%--55.62\% over SDF-only. This asymmetry does not imply that DG is generally superior to SDF. Rather, under the conditions considered here, DG already provides an effective representation of the overall geometry, whereas SDF supplements it with query-point-specific boundary information.

Taken together, Table~\ref{tab:joint_configurations} and Fig.~\ref{fig:joint_encoding_improvements} show that 15 of the 18 matched comparisons yield lower average errors than both corresponding single-encoding baselines, while all 18 outperform SDF-only. The remaining three configurations yield improvements of $-0.12\%$, $-0.46\%$, and $-0.56\%$ relative to DG-only. Given that each configuration was independently trained only three times, these small negative values are not interpreted as statistically significant performance degradation; rather, they indicate that joint encoding does not guarantee further improvement for every configuration. The different configuration rankings between the two internal-hole cases and the L-shaped bracket, together with the variation among DG--SDF combinations within each case, further demonstrate that their complementarity is both case- and configuration-dependent. Consequently, the preferred joint configuration cannot be determined simply by combining the individually optimal DG-only and SDF-only settings.

Considering the final prediction errors, incremental improvements, and cross-case performance, we selected all\_bound+boundary as the common configuration. It achieves near-minimum average errors for the two-hole plate and single-hole cantilever, the lowest average error of 8.47\% for the L-shaped bracket, and the largest mean improvement over DG-only across the three cases, at 8.10\%. Moreover, it provides positive improvements over both matched single-encoding baselines in all three cases.

\FloatBarrier

\section{Conclusions and Future Work}\label{sec:conclusion}

This study developed the Load- and Geometry-adaptive Deep Operator Network (LG-DON) for predicting structural response fields under varying geometries and boundary excitations. The framework combines discrete geometry inputs with separate encodings of distributed traction and prescribed displacement. Its DG module aggregates pointwise features into a sample-level geometry representation, without requiring a prescribed vector of shape parameters as the network input. Building on the scalar SDF input used in Geom-DeepONet, the boundary-distance module introduces separate channels for selected physical boundary classes. These channels retain query-dependent distances to distinct boundaries, supplementing the overall geometry information supplied by DG. Together, these components provide a representation of geometry and loading for structural response prediction.

LG-DON was validated on three two-dimensional linear elasticity problems: a plate with two irregular holes, a cantilever beam with one irregular hole, and an L-shaped bracket with a varying external geometry. The numerical results demonstrated its ability to reproduce displacement and von Mises stress distributions as internal or external boundaries and loading conditions varied. For the two-hole plate, LG-DON reduced the overall mean relative $L_2$ error from 12.17\% for modified MIONet to 7.41\%, primarily through improved von Mises stress prediction. Separate load-amplitude and geometry-parameter extrapolation tests on the cantilever beam further showed that the trained model retained the principal displacement-response patterns beyond the training ranges, although prediction accuracy depended on the loading scenario and response field.

The geometry-encoding ablations identified the benefits of joint encoding and the conditions under which these benefits arise. With DG disabled, boundary achieved lower average test errors than outer in all three cases, supporting the use of separate boundary-distance channels for the problems considered. The largest reduction occurred for the L-shaped bracket, from 29.91\% to 11.99\%. With SDF features disabled, restricting DG inputs to varying boundaries yielded the lowest errors across the three cases. However, the individually preferred representations did not always form the best joint configuration. Across the three cases and six DG--SDF combinations, joint encoding outperformed both matched single-encoding baselines in 15 of the 18 comparisons and outperformed SDF-only in all 18. These results support conditional complementarity between sample-level geometry features and query-dependent boundary distances. The all\_bound+boundary configuration achieved near-minimum prediction errors for the two internal-hole cases and the lowest error for the L-shaped bracket, while providing the largest mean relative error reduction across the three cases compared with its matched DG-only baseline, at 8.10\%.

The present validation is limited to two-dimensional, static, small-deformation linear elasticity and the geometry and loading families examined. Prediction errors remained concentrated near boundaries and regions with large response gradients and increased under the tested extrapolation settings. Moreover, the discretization study compared models trained separately at different node counts; it does not establish transfer between discretizations without retraining. The current supervised formulation also depends on FEM-generated response data, and the boundary-distance inputs do not themselves enforce mechanical equilibrium or boundary conditions.

Future work should first assess how boundary classification and point-set sampling affect predictive accuracy across broader geometry families. It should also include tests of a single trained model across different discretizations. Learning-curve studies are needed to quantify sample requirements and identify where additional geometry--load data provide the greatest benefit. Incorporating equilibrium constraints \cite{zhong2024lgdon30}, discrete stiffness relations \cite{ahmed2025lgdon31}, or energy and variational formulations \cite{wang2024lgdon32,eshaghi2025lgdon33} offers a further direction for investigating data efficiency and local response accuracy. Extensions to three-dimensional structures, nonlinear constitutive behavior, and time-dependent loading will require corresponding developments in the formulation and validation.




\bibliographystyle{elsarticle-num}

\bibliography{cas-refs}


\end{document}